\documentclass[pre,aps,twocolumn]{revtex4}
\usepackage{bm}
\usepackage{epsfig} 

\newcommand{\rhobar}{\overline{\rho}}
\newcommand{\chin}{\chi_{4\mathrm{n}}}
\newcommand{\diff}{\Omega}
\newcommand{\taup}{\tau_\mathrm{p}}
\newcommand{\ee}{\mathrm{e}}
\begin{document}

\title{Subdiffusive motion in kinetically constrained models}
\author{Robert L. Jack}
\affiliation{Department of Physics, University of Bath,
Bath BA2 7AY, UK}
\affiliation{Department of Chemistry, University of California
at Berkeley, Berkeley, CA 94720, USA}
\author{Peter Sollich}
\affiliation{King's College London, Department of Mathematics,
London WC2R 2LS, UK}
\author{Peter Mayer}
\affiliation{Department of Chemistry, Columbia University, 3000
Broadway, New York, NY~10027, USA}

\begin{abstract}
We discuss a kinetically
constrained model in which real-valued local densities fluctuate
in time, as introduced recently by Bertin, Bouchaud and Lequeux.  
We show how the phenomenology of this model can be reproduced
by an effective theory of mobility excitations propagating
in a disordered environment. 
Both excitations and probe particles have subdiffusive motion,
characterised by different exponents and operating on different time
scales.  We derive these exponents, 
showing that they depend continuously on one of the parameters of 
the model.
\end{abstract}


\maketitle

\section{Introduction}

There has been considerable recent interest 
in the hypothesis that glassy materials can be described
by coarse-grained models with simple thermodynamic
properties and non-trivial kinetic constraints 
\cite{RitortS03,GarrahanC02,JungGC04,BBL05,
BGC-Fick,Chandler06,knights,Berthier-chi4} 
These models capture the dynamically heterogeneous nature of 
glass-formers \cite{DHReviews}:
the implicit assumption is that microscopic details of
the glass-former are important only insofar as they
set the parameters of the coarse-grained dynamical theory.
Some kinetically constrained models describe particles
hopping on a lattice~\cite{KA,TLG}; in other cases, binary
(Ising) spins are used~\cite{GarrahanC02,FredricksonA84,EisingerJ93}, 
where the two states of the spin represent `mobile' and `immobile' 
regions of the liquid.  

In a recent paper, Bertin, Bouchaud and
Lequeux (BBL) \cite{BBL05} discussed a kinetically constrained model
in which molecular degrees of freedom
are modeled by a real-valued local density, defined on a lattice.  
Loosely speaking, regions of high density correspond to immobile
sites in the spin description of~\cite{FredricksonA84}, and regions of low 
density correspond to mobile sites.  However, the continuous
range of densities in the BBL model captures the fact that a glass-forming
system has a variety of local packings, which may not permit a 
simple decomposition into mobile and immobile.  The continuous
range of densities leads to a continuous range of mobilities, resulting
in a very broad distribution of relaxation times, characteristic
of glassy behaviour.

As discussed in Ref.~\cite{BBL05}, relaxation in the BBL
model occurs by means of  `mobility excitations' that
propagate subdiffusively across the system.  Links between
broadly distributed relaxation times and subdiffusive
motion of particles are quite familiar in theories of glass-forming
liquids~\cite{trap}.  Here, we focus initially on the
motion of mobility excitations, by coarse-graining
the BBL model onto an effective theory for these excitations.
This procedure represents a very simple example of the coarse-graining 
of glassy materials that was proposed by Garrahan and 
Chandler~\cite{GarrahanC02}.  The resulting effective theory is
a disordered generalisation of the one-spin facilitated
Fredrickson-Andersen (FA) model~\cite{FredricksonA84}.  The
possibility of subdiffusive motion in this model is
important when comparing model results with experiments 
on supercooled liquids: on approaching the experimental 
glass transition time scales increase dramatically, while the length 
scales associated with dynamical heterogeneity grow more 
slowly~\cite{DHReviews,Qiu03,Berthier-expt-long}.
In the dynamical facilitation picture~\cite{GarrahanC02},
the motion of mobility excitations leads
to a relation of the form $\tau\sim \xi^z/\diff$
where $z$ is a dynamical exponent,
$\tau$ the relaxation time scale, $\xi$ the length
scale associated with dynamical heterogeneity,
and $\diff$ a (possibly temperature dependent) 
constant~\cite{Whitelam04}.
Subdiffusion of these excitations corresponds to an exponent
$z>2$, consistent with a time scale that increases much
more quickly with the corresponding length scale than for ordinary diffusion.

In this paper we focus throughout on the one-dimensional case, where
subdiffusion effects are most pronounced~\cite{BBL05}. Analysis of the
disordered 
FA model leads us to two main results.  Firstly, we are able to
explain the scaling exponents observed in \cite{BBL05}.  In
particular, while the disorder in both the BBL and disordered FA
models is fluctuating, we explain why excitations propagate
with the scaling laws expected for a particle moving in a quenched
random environment.  Secondly, we consider the motion
of probe particles in the BBL and disordered FA models.  These
particles propagate subdiffusively, but with scaling laws that
are different from those of the mobility excitations.  

The form of the paper is as follows. In section~\ref{sec:models},
we define the BBL model and the disordered FA model.
In section~\ref{sec:chi4}, we use four-point correlation functions
\cite{Berthier-chi4,FranzDPG99,Berthier04,ToninelliWBBB05} 
to investigate the subdiffusive propagation of mobility
excitations, and we discuss the associated scaling exponents. In
section~\ref{sec:probes}, we consider the motion of probe particles
in the BBL model, and show that this behaviour can also be reproduced
in the disordered FA model.

\section{Models}
\label{sec:models}

\subsection{BBL model}

The (one-dimensional) BBL model is defined \cite{BBL05}
for a chain of continuous
densities $\{\rho_i\}$, constrained to $0<\rho_i<2$. Dynamical
moves involve rearrangement of the density between adjacent
pairs of sites:
\begin{equation}
\rho_i, \rho_{i+1} \to \rho_i', \rho_{i+1}', \quad \hbox{rate }
f_{i} P_\rho(\rho_i',\rho_{i+1}'|\rho_i+\rho_{i+1}),
\label{equ:bbl_rates}
\end{equation}
where $f_{i}=\Theta(2-\rho_i-\rho_{i+1})$
is the facilitation function for bond $i$, between sites $i$ and $i+1$.
Here, 
$\Theta(x)$ is the Heaviside step function, so a density rearragement
between two sites can occur only if the total density on those sites
is less than two.
The distribution of densities after the rearrangement is
\begin{equation}
P_\rho(\rho_i,\rho_{i+1}|R) = A (\rho_i \rho_{i+1})^{\mu-1}
\delta(R-\rho_i-\rho_{i+1}),
\label{equ:Prho_can}\end{equation}
where the delta function enforces volume conservation. 
The parameter $\mu>0$ was motivated in~\cite{BBL05} in terms
of an interaction between the particles of the model. If 
$\mu>1$ then the density after the rearrangment
tends to be distributed equally between sites $i$ and $i+1$;
if $\mu<1$ then the density is more likely to accumulate on just
one of the sites.
The coefficient $A=[R^{1-2\mu}\Gamma(2\mu)/\Gamma(\mu)^2]$ 
is determined by the requirement that
\begin{equation}
\int_0^2\!\mathrm{d}\rho_i\int_0^2\!\mathrm{d}\rho_{i+1}\,
P_\rho(\rho_i,\rho_{i+1}|R) = 1,
\label{equ:Prho_norm}
\end{equation}
which means that all facilitated bonds rearrange with unit rate.
[Here, $\Gamma(\mu)$ is the usual Gamma function.]

These dynamical rules respect detailed balance with respect
to a steady state distribution $P_\mathrm{stat,BBL}(\{\rho_i\})$
that factorises between sites.  In the
grand canonical ensemble we have
\begin{equation}
P_\mathrm{stat,BBL}(\{\rho_i\})=\prod_i P_\mathrm{s}(\rho_i),\qquad
P_\mathrm{s}(\rho) \propto \rho^{\mu-1} e^{\gamma\rho},
\label{equ:bbl_grand}
\end{equation}
normalised so that $
\int_0^2\!\mathrm{d}\rho\, P_\mathrm{s}(\rho)=1$.
Our notation differs from \cite{BBL05} in that we use $\gamma$
for the Lagrange multiplier conjugate to density, reserving
$\beta$ for the inverse temperature of the FA model.

It is clear from Eq.~(\ref{equ:bbl_rates}) that motion is
only possible across bonds with $f_i=1$.  We refer to these
as facilitated bonds.
The steady state contains a finite fraction of facilitated bonds,
which we denote by
\begin{equation}
\eta\equiv\langle f_i \rangle.
\label{equ:def_eta}
\end{equation}
We also define the mean density,
\begin{equation}
\rhobar \equiv \langle \rho_i \rangle.
\end{equation}

Facilitated bonds in the BBL model are the fundamental
mobility excitations
in the system.  The interesting scaling limit is the one of maximal mean
density, $\rhobar\to 2$, where facilitated bonds are rare ($\eta\ll
1$).  In this limit, $\gamma$ is large, and we have
\begin{eqnarray}
\rhobar &=& 2 - \gamma^{-1} + \mathcal{O}(\gamma^{-2}),  \\
\eta &=& 2\gamma \exp(-2\gamma) \frac{\Gamma(\mu)^2}{\Gamma(2\mu)}
[ 1 + \mathcal{O}(\gamma^{-1}) ]
\end{eqnarray}
consistent with~\cite{BBL05}.  
It was further observed in \cite{BBL05} that the dynamics 
of these excitations
in the BBL model can be represented schematically by the processes 
\begin{eqnarray}
01\leftrightarrow 11\leftrightarrow 10
\label{equ:eff_diff}
\end{eqnarray} 
where a 1 represents a facilitated bond ($f_i=1$ or $f_{i+1}=1$,
respectively), and a $0$ an unfacilitated bond ($f_i=0$ or $f_{i+1}=0$).  
The above two-step process then produces effective diffusion of
excitations. When excitations meet, they can coagulate via e.g.\ $101\to
111\to 011\to 010$; running through the steps in reverse, a single
excitation can also branch into two. Excitations can never be created
unless there is already an excitation present on a neighbouring bond,
and this is the key motivation for the effective FA models presented
below.

When excitations are rare, the rate-limiting step in the effective
diffusion is the creation of a new excitation, $01\to 11$. To obtain
the typical rate for this process, consider a density rearrangement event
across bond $i+1$:
\begin{eqnarray}
\rho_i, \rho_{i+1}, \rho_{i+2} \to \rho_i, \rho_{i+1}', \rho_{i+2}'.
\label{equ:AtoAA}
\end{eqnarray}
The process $01\to 11$ occurs when
the second bond is facilitated in both initial and final
states, while the first bond is
facilitated only in the final state.  That is,
\begin{eqnarray}
\nonumber
\rho_{i+1}+\rho_{i+2} = \rho_{i+1}' + \rho_{i+2}' &<& 2, \\
\nonumber
\rho_{i}+\rho_{i+1} &>& 2, \\ 
\rho_{i}+\rho_{i+1}' &<& 2.
\end{eqnarray}
To work out the typical rate with which these processes occur, we
should perform a steady-state average over all initial configurations
with the prescribed mobility 
configuration $(f_i=0,f_{i+1}=1)$, corresponding to the first two
conditions listed. In addition, however, we condition
on $\rho_i$, which strongly influences the rate if it is close to
2: the third condition given above can then only be met if
$\rho_{i+1}'$ is very small. Thus, we consider the average rate for the process
$(\rho_i=2-\epsilon_i,f_i=0,f_{i+1}=1) \to
(\rho_i=2-\epsilon_i,f_i=1,f_{i+1}=1)$, which can only occur via a density
rearrangement across bond $i+1$ as written above.  The
steady-state distribution factorises between sites and so we have for
this rate, denoted by $r_i(\rho_i)$:
\begin{widetext}
\begin{equation}
 r_i(\rho_i)=\frac{\int_{\epsilon_i}^{2}\!\mathrm{d}R 
  \int_{\epsilon_i}^R \!\!\mathrm{d}\rho_{i+1}
\,
 P_\mathrm{s}(\rho_{i+1})
P_\mathrm{s}(R-\rho_{i+1}) 
\int_0^{\epsilon_i}
\mathrm{d}\rho'_{i+1}\,
\int_0^2\mathrm{d}\rho_{i+2}'P_\rho(\rho_{i+1}',\rho_{i+2}'|R)}
{\int_{\epsilon_i}^{2}\!\mathrm{d}R 
  \int_{\epsilon_i}^R \!\!\mathrm{d}\rho_{i+1}\, 
 P_\mathrm{s}(\rho_{i+1})
P_\mathrm{s}(R-\rho_{i+1}) }
\label{equ:rate_epsilon}
\end{equation}
\end{widetext}
where we have introduced 
$R=\rho_{i+1}+\rho_{i+2}$. The integral over $\rho_{i+2}'$ in the
numerator gives $A[\rho_{i+1}'(R-\rho_{i+1}')]^{\mu-1}$, and the one
over $\rho_{i+1}'$ then a normalized incomplete Beta function
$B(\mu,\mu;\epsilon_i/R)/B(\mu,\mu)$. The remaining average over $R$
(and $\rho_{i+1}$) becomes concentrated around $R=2$ for large
$\gamma$, so that in this limit
\begin{equation}
r_i(\epsilon_i)=\frac{B(\mu,\mu;\epsilon_i/2)}{B(\mu,\mu)}
=\frac{\int_0^{\epsilon_i/2}dv\,v^{\mu-1}(1-v)^{\mu-1}}
{\int_0^1 dv\,v^{\mu-1}(1-v)^{\mu-1}}
\end{equation}
Recalling that the local density is $\rho_i=2-\epsilon_i$, we note that
dense sites (those with small $\epsilon_i$) lead to small rates
$r_i$.

The relaxation of the BBL model on long time scales is determined
by sites with small $r_i$.  For this reason, it is convenient to
deduce the distribution of this rate from that of $\rho_i$, or
equivalently $\epsilon_i$. From Eq.~(\ref{equ:bbl_grand}) one sees for
large $\gamma$ that the variation of the power law factor
$\rho^{\mu-1}$ near $\rho=2$ can be neglected, so that
$P_\mathrm{s}(\epsilon_i)=\gamma\exp(-\gamma\epsilon_i)$. The typical
values of $\epsilon_i$ are therefore small, $\epsilon_i\sim
\gamma^{-1}$, and we can expand the rate as
\begin{equation}
r_i(\rho_i)\simeq a\epsilon_i^\mu, \qquad a=\frac{\Gamma(2\mu)}
{\mu\, 2^\mu \Gamma(\mu)^2}
\label{equ:rate_expansion}
\end{equation}
Transforming then from the distribution of $\epsilon_i$ to $r_i$ gives
\begin{eqnarray}
P_{\mathrm{s},r}(r) &=& 
  \left[ P_\mathrm{s}(\epsilon_i) 
       \left|\frac{\mathrm{d}r_i(\epsilon_i)}{\mathrm{d}\epsilon_i}\right|^{-1}
 \right]_{r=r_i(\epsilon_i)} \nonumber \\
  &=&\frac{1}{\mu\diff_0} (r/\diff_0)^{(1/\mu)-1} \ee^{-(r/\diff_0)^{1/\mu}}
\label{equ:rate_dist}
\end{eqnarray}
where
\begin{equation}
\diff_0 =a \gamma^{-\mu} \sim \gamma^{-\mu}
\end{equation}
is a microscopic rate, which acts as an upper cutoff on the 
distribution of rates. [The notation $u \sim v$ means that
$u$ and $v$ are proportional to each other in the relevant limit
(large $\gamma$).] It is important to note that the small-$r$ scaling
of the distribution $P_{\mathrm{s},r}(r)\sim r^{(1/\mu)-1}$, which we
derived above in the limit $\gamma\to\infty$, also applies at finite
$\gamma$. This is because the rate for small $\epsilon_i$ always
scales as in Eq.~(\ref{equ:rate_expansion}), and the probability
density $P_\mathrm{s}(\epsilon_i)$ of $\epsilon_i$ approaches a
constant for small $\epsilon_i$ for all $\gamma$.

The long-time behaviour of the BBL model is now controlled by
the behaviour of $P_{\mathrm{s},r}(r)$
at small $r$, and, in particular, by the exponent $\mu$. The time
for a mobility excitation to diffuse across bond $i$ is of order
$1/r_i$. 
The average diffusion time $\langle 1/r\rangle$, with the
average taken over the distribution $P_{\mathrm{s},r}(r)$, then shows a
change of behaviour at $\mu=1$: for $\mu<1$ it is finite, while for
$\mu>1$ it diverges. This motivates why subdiffusion occurs in the
second case: for arbitrarily long times $t$ there are a significant
number of barriers to mobility diffusion that have transmission rate $<1/t$.

\subsection{Effective FA model}

We now describe the effective model that captures
the dynamics of the mobile bonds 
on large length and time scales. 
In this model, the bonds of the BBL model are represented
by a chain of binary variables $\{n_i\}$, where $n_i=1$ if
the bond between sites $i$ and $i+1$
of the BBL model is mobile, and $n_i=0$ otherwise. The variable
$n_i$ corresponds to the BBL variable $f_i$. The process
of (\ref{equ:eff_diff}) is then
\begin{eqnarray}
(n_i=0, n_{i+1}=1) \to (n_i=1, n_{i+1}=1)
.
\end{eqnarray} 
It occurs with a rate, $r_i \ee^{-\beta}$, and  the
reverse process occurs with rate $r_i$.  Here, $\ee^{-\beta}$
determines the concentration of sites with $n_i=1$, while $r_i$
is a site-dependent rate whose fluctuations
capture the effect of the fluctuating density $\rho_i$ in the
BBL model. In our effective model we use the
convention $0<r_i<1$; taking a maximal rate of unity
sets the unit of time.  To mimic
the distribution of rates in the BBL model,  we define the
disordered FA model so that
the $r_i$ are distributed independently in the steady state,
with 
\begin{equation}
P_r(r_i)=(1/\mu)r_i^{(1/\mu)-1}, \qquad r_i<1
\label{equ:P_r}
\end{equation}
in accordance with (\ref{equ:rate_dist}).

The rate $r_i$ in the disordered FA model reflects the local density in
the BBL model: it is a fluctuating variable.  In the
dynamics of the disordered FA model, we account for this fact by
randomising $r_i$ when the process corresponding to (\ref{equ:AtoAA})
occurs.  Hence, we define 
our disordered FA model by the dynamical rules:
\begin{eqnarray*}
(n_{i}, n_{i+1}, r_{i}) &\to& (n_{i}, 1-n_{i+1}, 
r_{i}'), \nonumber
\\ & & \quad \hbox{rate } n_{i} r_i
e^{\beta(n_{i+1}-1)}
P_\mathrm{ann}(r_{i}') 
  \nonumber\\
(n_{i}, n_{i+1}, r_{i}) &\to& (1-n_{i}, n_{i+1}, r_{i}'), 
\nonumber \\ & &
\quad \hbox{rate } n_{i+1} r_i
e^{ \beta(n_{i}-1)}
 P_\mathrm{ann}(r_{i}'), 
  \nonumber
\end{eqnarray*}
where 
\begin{equation}
P_\mathrm{ann}(r)\propto rP_r(r),
\end{equation}
(explicitly, $P_\mathrm{ann}(r)=[(1/\mu)+1] r^{1/\mu}$), 
and the variables $n_i\in\{0,1\}$ and $0<r_i<1$ reside on the sites
and bonds of the FA lattice respectively.
We identify this model as a disordered variant of the
FA model \cite{FredricksonA84}, since the case $P_r(r)=\delta(r-1)$
is the one-spin facilitated one-dimensional FA model.
We refer to it as the bond-disordered FA model since the rates $r_i$
are associated with the bonds of the (FA) lattice.

The `annealed' distribution of rates after randomisation,
$P_\mathrm{ann}(r_{i}')$, is constructed such that the model obeys
detailed balance with respect to 
\begin{equation}
P_\mathrm{stat,FA}(\{n_i\},\{r_i\}) \propto \prod_i P_r(r_i) e^{-\beta n_i}
.
\end{equation}
The stationary density of sites with $n_i=1$ is 
\begin{equation}
c\equiv \langle n_i \rangle = (1+e^\beta)^{-1},
\end{equation}
which plays the part of the parameter $\eta$ defined in (\ref{equ:def_eta}).
The only other parameter in the model is $\mu$, which corresponds directly
with $\mu$ in the BBL model. 

Several other comments are in order. We chose $0<r<1$ above, for
convenience. As a result, we do not expect direct correspondence between
time units in the original and effective models [comparing
(\ref{equ:rate_dist}) and~(\ref{equ:P_r}), we have effectively set
the prefactor $\diff_0$ to unity].
There is also no exact correspondence between the
steady states: in the FA model, there are no spatial correlations at
all between the $n_i$, whereas in the original BBL model neighbouring
excitations $(f_i,f_{i+1})$ are correlated via the density variable
$\rho_{i+1}$. Finally, also the way rates $r_i$ are linked to creation
and destruction of excitations
does not match exactly. In the original BBL model, we saw above that
the process $(f_i=0, f_{i+1}=1) \to (f_i=1, f_{i+1}=1)$ is controlled
by the density $\rho_i$, and is slow when $\rho_i$ is close to
2. Translating to the dual lattice of the FA model, this corresponds
to the controlling rate for $(n_i=0, n_{i+1}=1) \to (n_i=1,
n_{i+1}=1)$ being associated with the bond between $n_{i-1}$ and
$n_i$, {\em not} with the bond between $n_i$ and $n_{i+1}$ as we have
posited. Thus e.g.\ the
transient appearance of an excitation, $01\to 11\to 01$,
randomizes $r_i$ in our FA model but does not change $\rho_i$ in the
BBL model so that $r_i$ remains unchanged as well.
On the other hand, in an effective diffusion step
$01\to 11\to 10$ in the BBL model, the second step involves a
rearrangement across bond $i$ and so a randomisation of $\rho_i$ and
hence $r_i$. This is correctly captured in the FA model, and as
effective diffusion is the key process in the dynamics we expect our
model to give a qualitatively correct description of the BBL dynamics.

\subsection{Model variants}

\subsubsection{Grand canonical BBL model}

The grand canonical expression
(\ref{equ:bbl_grand})
motivates us to define a modified BBL model in which volume
is not conserved.  We use the
same dynamical rule (\ref{equ:bbl_rates}), but replace 
$P_\rho(\rho_i,\rho_{i+1}|R)$ by
\begin{equation}
P_\rho'(\rho_i,\rho_{i+1}) = A' (\rho_i \rho_{i+1})^{\mu-1}
e^{\gamma(\rho_i+\rho_{i+1})} \Theta(2-\rho_i-\rho_{i+1})
\label{equ:Prho_grand}
\end{equation}
where the final state is now independent of the volume in the initial
state.  These dynamical rules preserve the same equilibrium distribution
as that of the original BBL model, as given in~(\ref{equ:bbl_grand}).
The constant of proportionality $A'$ is set by
$\int_0^2\!\mathrm{d}\rho\int_0^2\mathrm{d}\rho'\,P_\rho'(\rho,\rho')=1$
so that bonds rearrange with unit rate, as in the original model.

We will find that propagation of mobile bonds is similar in models
with and without conserved density, although the relaxation of 
density fluctuations will clearly be different.

\subsubsection{Site-disordered FA models}

We also define a site-disordered FA model, in which we associate
random rates $r_i$ with the sites of the FA chain, instead of the bonds.
The dynamical rules are then
\begin{eqnarray}
(n_i,n_{i+1},r_i,r_{i+1}) &\to& 
(n_i,1-n_{i+1},r_i,r'_{i+1}), 
\nonumber \\ & &
\quad
\hbox{rate } n_{i} r_{i+1} e^{\beta(n_{i+1}-1)} P_\mathrm{ann}(r'_{i+1}) 
\nonumber
\\
(n_i,n_{i+1},r_i,r_{i+1}) &\to& 
(1-n_i,n_{i+1},r'_i,r_{i+1}), 
\nonumber \\ & &
\quad
\hbox{rate } n_{i+1} r_i e^{\beta(n_{i}-1)}
P_\mathrm{ann}(r'_i)
\nonumber
\end{eqnarray}
As for the bond-disordered FA model, also this model does not exactly
capture how rates are linked to rearrangements in the BBL model; but
it does provide for rates to be randomised every time an excitation
makes an effective diffusion step, which is the key property for the
physics. Indeed, we will see below that the excitations behave
similarly for bond and site disorder. However,
on introducing probe particles to these disordered FA models,
one finds that the site-disordered model provides a better match to
the BBL model dynamics. The reasons for this will be explained below.

\subsubsection{FA models with quenched disorder}

Finally, it is convenient to define FA models with quenched
disorder, in which the
rates $r_i$ do not depend on time. The distribution of rates
is simply $P_r(r_i)$ in that case. Interestingly, we will find that 
quenching the disorder in this way has very little effect on dynamical
correlations (after disorder averaging).
We note that the quenched bond-disordered FA model has a mapping
to a disordered model of appearing and annihilating defects (AA model),
and inherits from the latter an exact duality mapping, as in the pure
case~\cite{JMS06}.

\section{Mobility excitations}
\label{sec:chi4}

\begin{figure*}
\epsfig{file=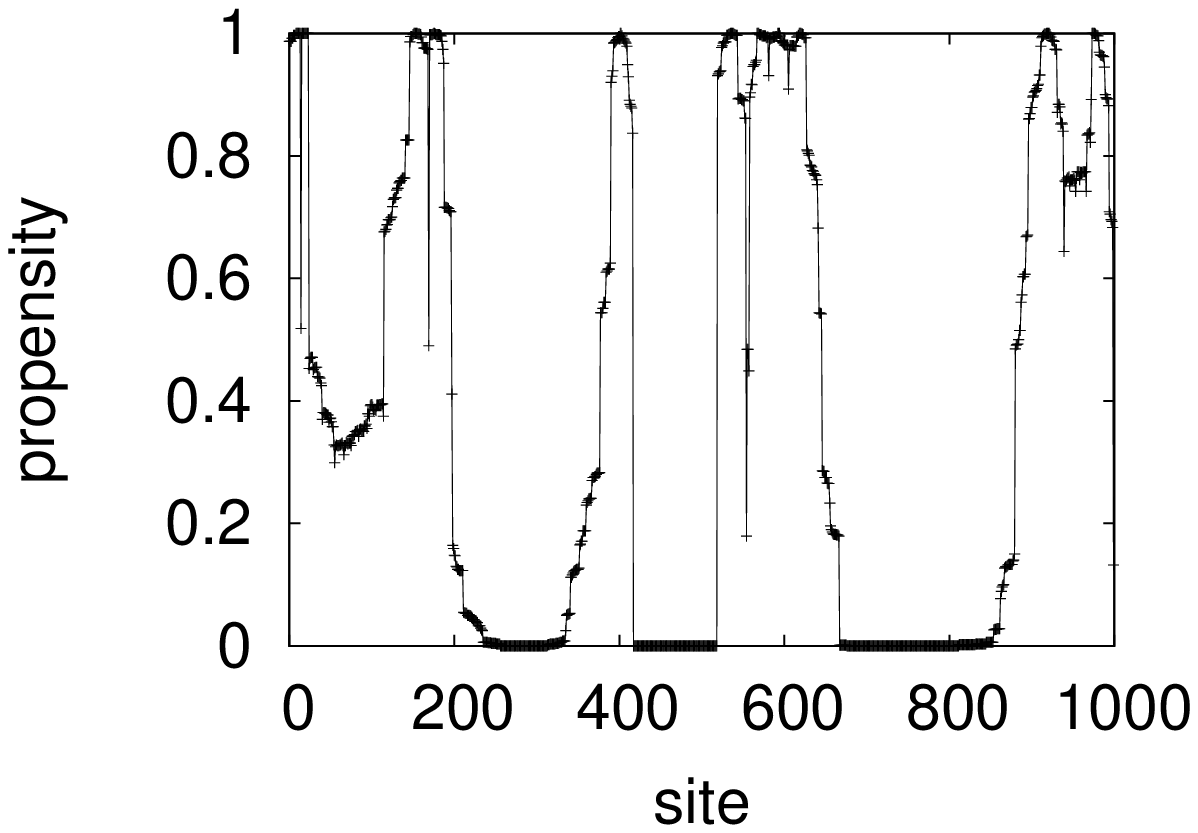,width=5.1cm}
\epsfig{file=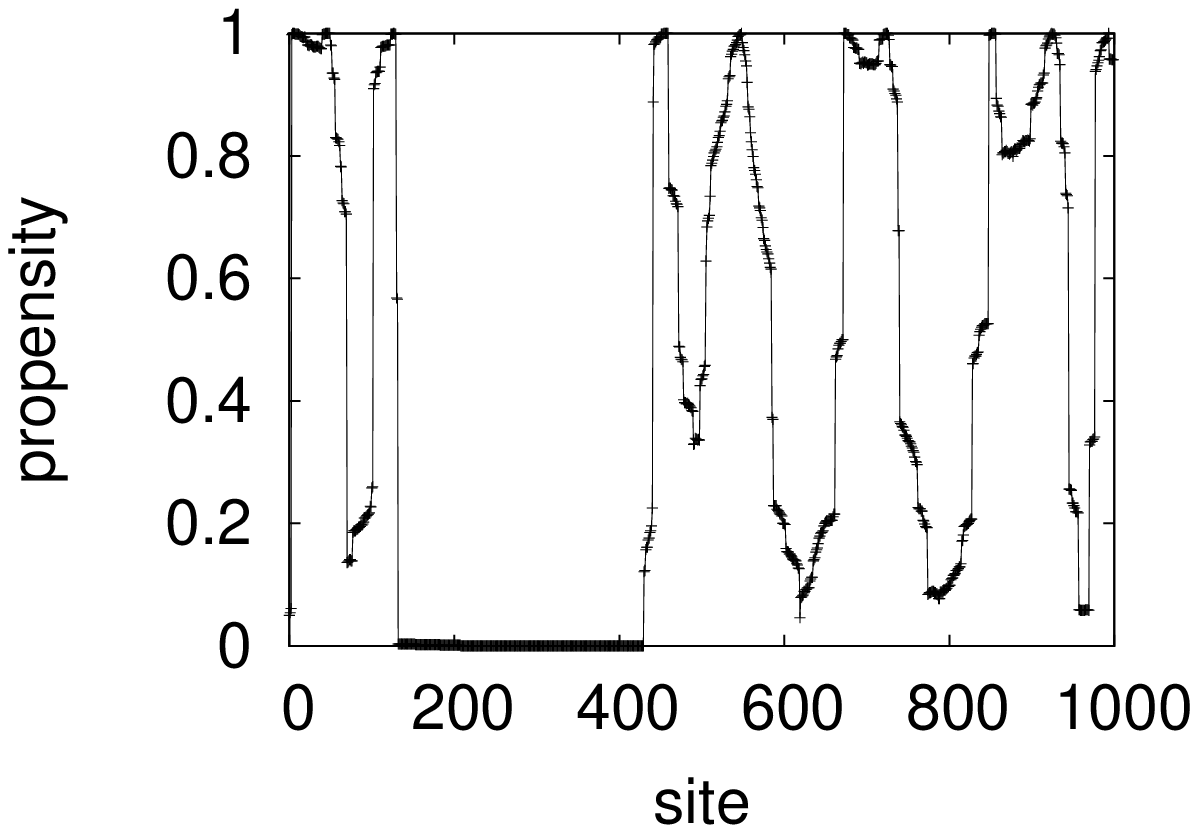,width=5.1cm}
\epsfig{file=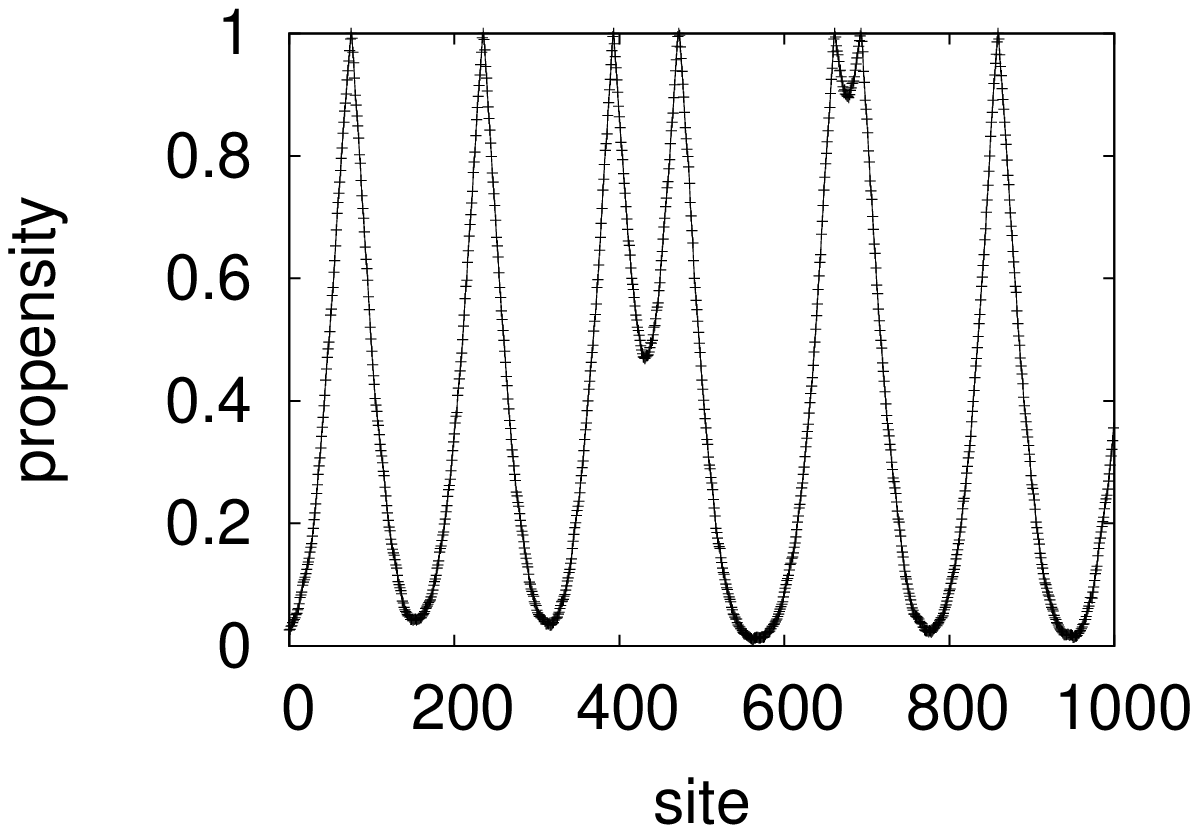,width=5.1cm}
\caption{Typical realisations of the propensity, with times
such that the spatially averaged persistence function
satisfies $1-\langle p(t)\rangle\simeq 0.4$. 
Large values
of the propensity indicate sites that are very likely
to have relaxed, on this time scale.  The models with
subdiffusive dynamics 
have large jumps in the propensity, which arise from sites
that relax very infrequently.
(Left) BBL model at $\mu=2$ and $\eta=0.01$. 
(Center) Bond-disordered FA model at $\mu=2$ and $c=0.01$.
(Right) Pure FA model at $c=0.01$. In the pure case, there 
are no barriers for excitation propagation, 
and the propensity is smooth [except near sites with $n_i(t=0)=1$,
where it is maximal]. 
}
\label{fig:prop}
\end{figure*}

We now consider the dynamics of the mobility excitations in the BBL
model, always in the interesting limit where $\rhobar$ is close to two.
First consider the regime where the parameter $\mu$ is small (much less
than unity). The BBL model in its steady state then has a bimodal
distribution of densities with sharp peaks near zero and two. In that case,
it describes diffusing vacancies in a one-dimensional solid (i.e.\
high-density background).  The
disordered FA models, on the other hand, all reduce to the pure FA
model in the limit of small $\mu$.
All models then exhibit dynamical scaling when the excitation
density is small, with exponents
\begin{equation}
(z,\nu) = (2,1), \qquad \mu<1
\label{equ:exponents_pure}
\end{equation}
Here $z$ is the dynamical exponent that sets the relative
scaling of space and time, while the correlation length scales as the
average distance between excitations, i.e.\
$\xi\sim\eta^{-\nu}$ for the BBL model and $\xi\sim c^{-\nu}$
for the FA case. 

However, the case of $\mu>1$ is qualitatively
different from that of small $\mu$.  For example, 
the mean time associated with rearrangements is $\langle 1/r \rangle$ which
diverges for $\mu>1$ as explained above [recall equations (\ref{equ:rate_dist})
and (\ref{equ:P_r})]. 
We therefore expect the disorder to have a strong effect: 
this is clear from plots of the
propensity~\cite{HarrowellPropensity}, which we define in terms of
the persistence function, $p_i(t)$.  This function takes a value
of unity if the state of site $i$ has not changed between time zero and 
time $t$, and $p_i(t)=0$ otherwise. The (time dependent) 
propensity for a given initial condition
of the system is then $1-\langle p_i(t) \rangle_\mathrm{dyn}$, where
the average is over the stochastic dynamics of the
system, but with the initial condition fixed~\cite{HarrowellPropensity}.
We show sample plots in figure~\ref{fig:prop}. In the BBL model,
sites with density close to two act as barriers to propagation
of mobility; in the FA model the same effect arises from bonds
with small rates.

A more quantitative measure of the effect of the disorder
is its effect on dynamical scaling. It was observed
in \cite{BBL05} that the BBL model has scaling exponents close to
\begin{equation}
(z,\nu) = (1+\mu,1), \qquad \mu>1.
\label{equ:exponents_dis}
\end{equation}
An effective model of a single excitation propagating in a
quenched environment of random energy barriers gives this scaling,
if the distribution of rates for crossing the barriers is 
$P_r(r)$ \cite{BouchaudG90}.
We will discuss below why this quenched result is applicable
to the BBL model, which has no quenched disorder. First, though, we
show that this subdiffusive scaling can be observed in the four-point
functions of both BBL and disordered FA models.

We consider the correlation function
\begin{equation}
G_4(x,t)=\langle \delta p_{i+x}(t) \delta p_i(0)\rangle
\end{equation}
where $\delta p_i(t) = p_i(t) - \langle p_i(t) \rangle$; $p_i(t)$
is the persistence operator defined above, and the averages are
over both initial conditions and the stochastic dynamics.
The normalised four point susceptibility is
\begin{equation}
\chin(t) \equiv \sum_x G_4(x,t) / G_4(0,t)
\end{equation}
In one dimension $\chin(t)$ is a direct measurement
of a growing length scale, when normalised in this way
\cite{ToninelliWBBB05,JackBG05}. 
(Note that we evaluate averages in an ensemble with
fixed `chemical potential' $\gamma$, so that the mean
density $\rhobar$ is allowed to fluctuate.  Since
the dynamics conserve $\rhobar$, this choice does affect
the value of $\chin(t)$~\cite{Berthier-chi4}.)
In the scaling limit ($\rhobar\to2$
from below),
we expect
\begin{equation}
\chin(t) \sim (\diff t)^{1/z} f( \diff t/\xi^z )
\label{equ:chi4_scaling}
\end{equation}
where $\xi$ is the correlation length whose scaling was given above,
$z$ is given by (\ref{equ:exponents_pure}) or
(\ref{equ:exponents_dis}), as appropriate, $\diff$
is a microscopic rate and $f(x)$ is a scaling
function that is constant at small $x$ and decreases as $x^{-1/z}$ for
large $x$. 
We argue below that the for the BBL model, the
rate $\diff$ is equal to $\diff_0$ [recall (\ref{equ:rate_dist})],
while for the FA model, we have $\diff\sim c$ (for small excitation
density $c$).  

\begin{figure} 
\epsfig{file=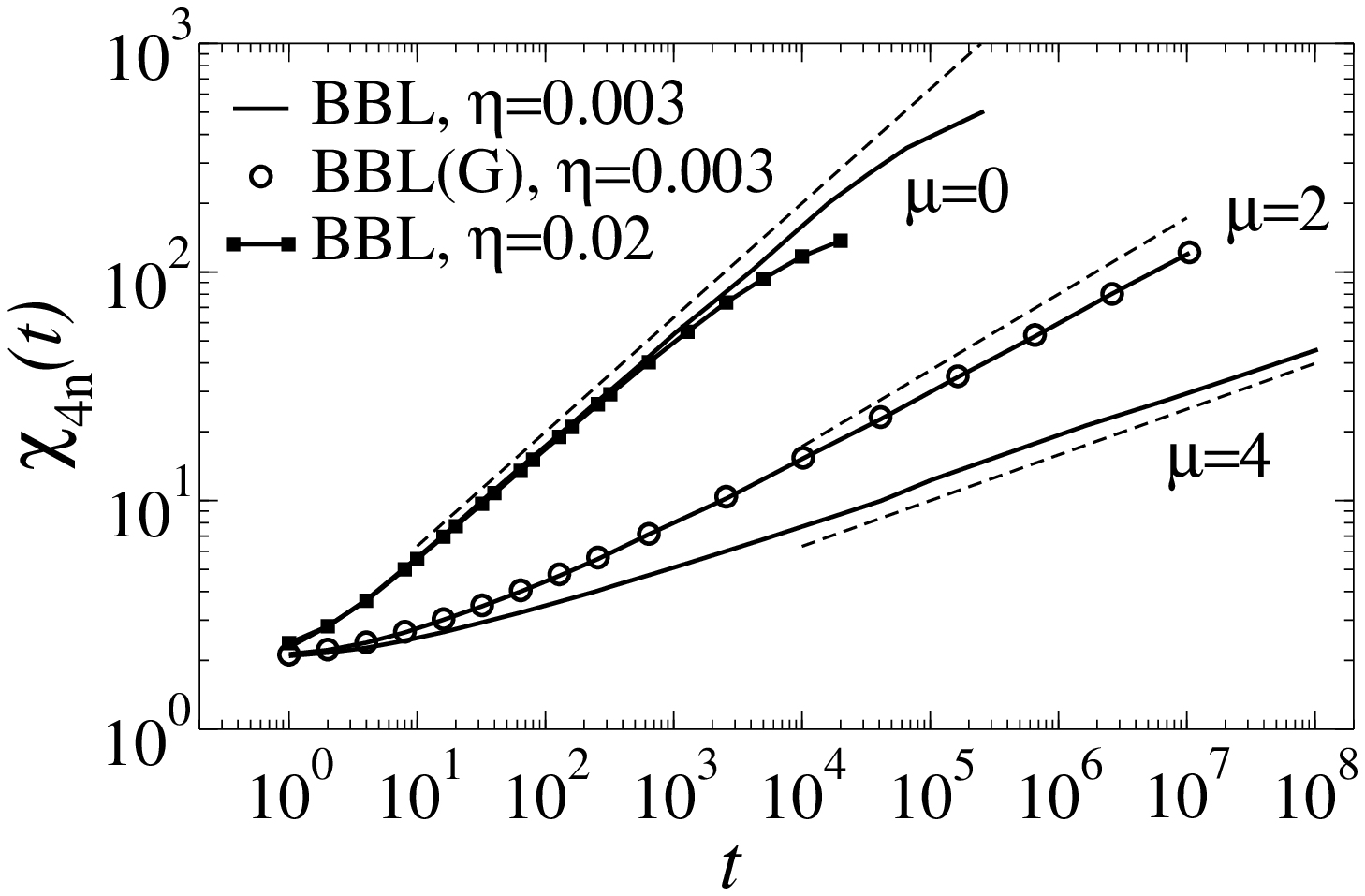,width=7.7cm}
\epsfig{file=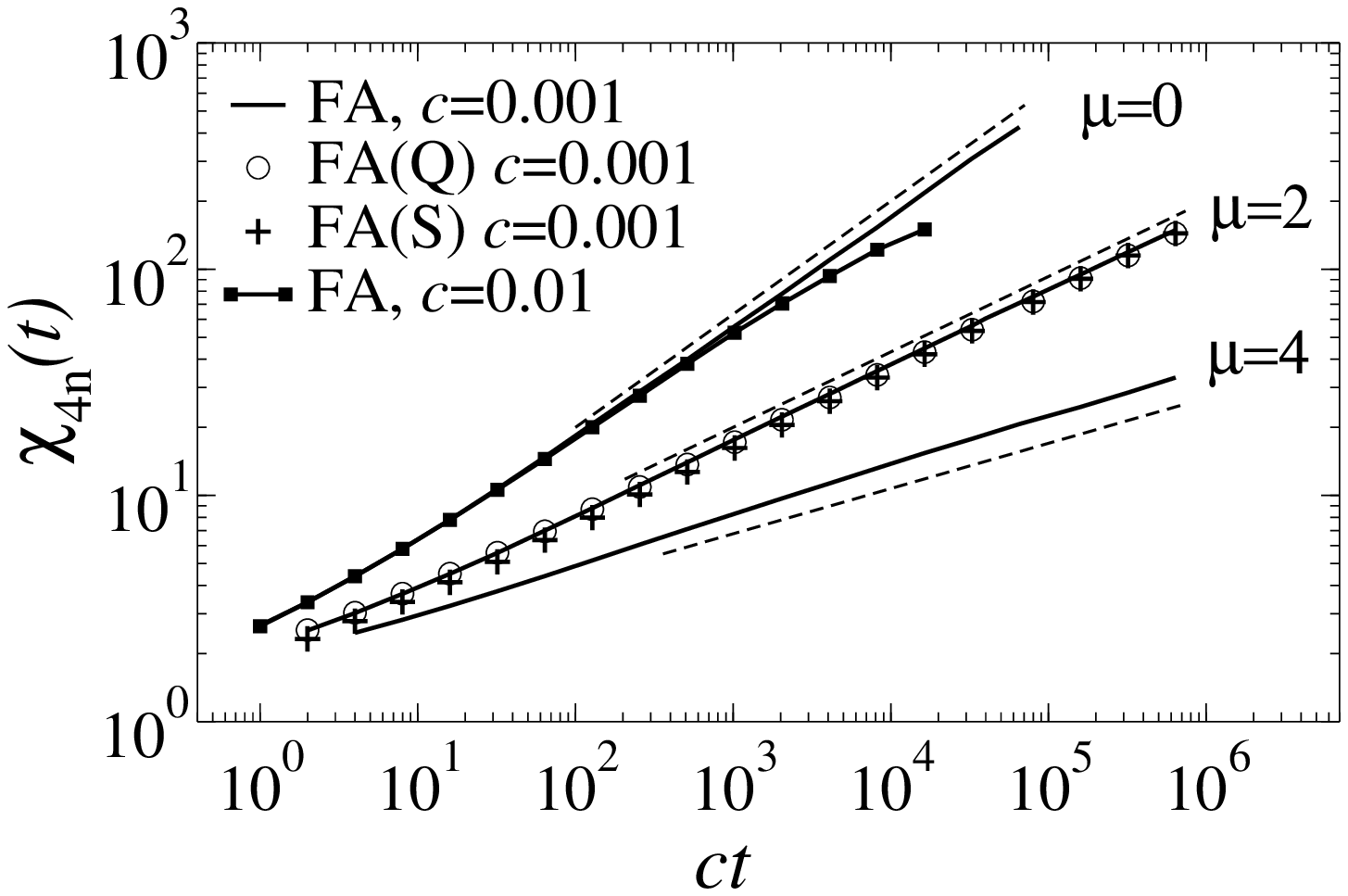,width=7.7cm}
\caption{Subdiffusion in the FA and BBL models is apparent in the four
point susceptibility. (Top) We show $\chin(t)$ in the BBL model
at various $\mu$ and $\eta$. For the case $\mu=0$, we use a binary
distribution of on-site densities 
$P_\mathrm{s}(\rho)\propto\alpha\delta(\rho) + \delta(2-\rho-0^+)$, so
that $\eta=\alpha(2+\alpha)/(1+\alpha)^2$; the small offset $0^+$ ensures that
the facilitation constraint is well defined.  The dashed
lines show the power law predictions of
(\ref{equ:exponents_dis}), and simple diffusive scaling
(\ref{equ:exponents_pure}) for $\mu=0$. At 
large times, $\chi_{4\mathrm{n}}(t)$ saturates at a value of order 
$\eta^{-1}$: this is apparent in the data for $\mu=0$.  The 
label (G) denotes data for the grand canonical variant of the BBL model. 
(Bottom) Bond-disordered FA model data with the same values of $\mu$, 
showing that this effective model captures the four point correlations of
the BBL model. The labels (Q) and (S) denote data from the model with
quenched bond disorder and fluctuating site disorder respectively; the
different variants have very similar behaviour.}
\label{fig:chi4}
\end{figure}

We show results in figure~\ref{fig:chi4}.  Both models 
are consistent with (\ref{equ:chi4_scaling}); we also
find that the FA model
exhibits the same scaling as the BBL model, if we identify the
excitation densities $c$ and $\eta$. Hence we argue that the disordered
FA model is an appropriate effective theory for the BBL model.


Figure~\ref{fig:chi4} demonstrates further that BBL models with and
without conserved density behave very
similarly. We conclude that the conservation of density is not relevant
for scaling: this is consistent with the use of the disordered FA model
as an effective theory, since that model has no conserved density.
Quenching the disorder in the FA model has only very weak effects on
disorder-averaged properties such as $\chin(t)$; finally, differences
between bond-disordered and site-disordered models are also very small.

\subsection{Effective barrier models for single excitations}
\label{sec:eff}

Figure~\ref{fig:chi4} is clear evidence that both BBL and disordered
FA models have excitations that propagate subdiffusively at large
$\mu$. Further, the dynamical exponents for all the models
with $\mu>1$ seem to satisfy (\ref{equ:exponents_dis}).

For the FA model with quenched bond disorder, this result is
to be expected since the motion of independent random walkers
in this kind of environment is well-understood~\cite{BouchaudG90} 
and does indeed satisfy
(\ref{equ:exponents_dis}). However, it was argued in
\cite{BBL05} that fluctuating disorder should lead to
\begin{equation} \label{equ:exponents_ann}
z=\mu,\qquad \mu>2
\end{equation}
and $z=2$ otherwise. This result is inconsistent with the data.

We are not aware of any analysis
of fluctuating disorder that is slaved to the motion of the random
walker.  In this section, we give an argument that explains
the applicability of (\ref{equ:exponents_dis}) to the FA model
with fluctuating disorder, and hence to the BBL model.  While
this is not a rigorous proof, the various stages of the
argument have been verified by direct simulation.

To describe the motion of a single excitation
in a disordered environment, we consider a simple barrier 
model~\cite{BouchaudG90}.
A single particle moves on a chain of sites,
with independent random hop rates $\{r_i\}$ on the bonds, distributed
according to $P_r(r)$.  We consider both 
quenched and fluctuating disorder: if the disorder is fluctuating, 
then each random rate is redrawn from the distribution
$P_\mathrm{ann}(r)$ when the bond is traversed by the
random walker.  We have verified by simulation that both variants
of this model do indeed satisfy (\ref{equ:exponents_dis}).  We
explain this result using an argument related to that of 
le Doussal, Monthus and Fisher \cite{DoussalMF99}.  The 
effective dynamics scheme that we use
for the barrier model with quenched disorder was described
in~\cite{Jack-duality-08}, where it was shown that
the effective dynamics are a good description of the quenched
barrier model, as long as the exponent $\mu$ is large.
We give a brief description of the effective dynamics here,
referring to~\cite{Jack-duality-08} for details.

In \cite{DoussalMF99}, the authors proposed an effective dynamics
for a random walker in a (quenched) one-dimensional energy 
landscape, made up of `barriers' and `valleys'.  
At each stage of the effective dynamics,
the smallest barrier in the system is removed, and the particle moves
to the bottom of the valley that contains the origin. The
time associated with this process is
the inverse transmission rate of the barrier
that was removed. For models in which
the energy landscape has short-ranged correlations,
this effective dynamics mimics the real dynamics of the random walker.

For the quenched barrier model, every site is at zero energy,
and they are separated by barriers of varying heights.
The effective dynamics involves successive removal
of the smallest barriers.
Thus, at a given stage of the dynamics, the remaining
barriers divide the system into `effective traps'.  
As discussed in~\cite{Jack-duality-08}, the barrier
model requires a modification to the scheme 
of~\cite{DoussalMF99}, in that the time at which
barrier $i$ is removed depends both on the rate $r_i$
and on the widths of the effective traps to the left and right
of barrier $i$.  If the widths of these traps are $l_1$
and $l_2$, the time $\tau_i$ associated with barrier $i$ is
determined by $\tau_i^{-1}=r_i(1/l_1 +1/l_2)$.

To arrive at the subdiffusive scaling of the quenched barrier
model, we assume that, at time $t$, effective traps have
a typical width $\ell(t)$.  All barriers
with $\tau_i<t$ have been removed; typically, these
barriers have $r_i\gtrsim \ell(t)/t$~\cite{typical-foot}.  
Thus, the density of remaining barriers is 
\begin{equation}
\ell(t)^{-1} \simeq \int_0^{\ell(t)/t} \mathrm{d}r\, P_r(r)
\end{equation}
which yields (for large $t$)
\begin{equation}
t \sim \ell(t)^{1+\mu}.
\label{equ:t-ell}
\end{equation}
The mean square displacement of the diffusing
excitation scales with $\ell(t)^2$, 
so we identify the dynamic exponent $z=1+\mu$, consistent
with (\ref{equ:exponents_dis}).  Recall that the effective
dynamics scheme applies only for large $\mu$, so we
do not recover the diffusive result, $z=2$ for $\mu<1$.

We now apply this scheme to models with fluctuating
disorder.  The fluctuations in the disorder have two
main effects.  Firstly, once large barriers have been crossed, 
their rates are randomised. Thus, if multiple crossings
of the same large barrier are important for the quenched model,
we expect different behaviour for fluctuating disorder.  
However, a central assumption of the effective dynamics is that the time
to travel a distance $\ell$ is dominated by the
time required for the first crossing of the largest barrier between
the initial and final sites~\cite{Jack-duality-08}.  
Hence, multiple crossings of large barriers are ignored
in the effective dynamics, which should therefore
be consistent with fluctuating disorder.  
The second effect of fluctuating disorder is that
barrier transmission rates are being randomised as the 
excitation moves around, so a barrier which previously had
a large transmission rate may acquire a new rate that is very small.
This new rate would then act as a high barrier and so have a strong
effect on the resulting motion.  As before,
the time taken to move a distance $\ell$ will be given by the 
time taken to cross the largest barrier between initial and final states: 
this might be a barrier
that was present initially, or one that appeared as the excitation
moved through the system.  
The key point here is that the system is in a steady state, 
so the introduction of new barriers occurs with the same rate 
as the removal of barriers of the same size.
Since barriers are removed only when they are crossed, 
barriers that appear in the system are typically
of a size comparable with those that have already been crossed at
least once.  In the language of the effective dynamics, 
these barriers are `irrelevant'.  For these reasons, the effective dynamics
apply equally well to models with quenched and fluctuating disorder,
and we expect the dynamical exponent $z=1+\mu$ for both cases.

Of course, the situation would be very different if the fluctuating
disorder was annealed in a two-sided way, where every time an
excitation moved to a new site one randomises
the rates for both of the barriers adjacent to that site.
This would produce a continuous-time random walk~\cite{BouchaudG90},
with subdiffusion exponents as in Eq.~(\ref{equ:exponents_ann}).

To obtain the scaling of length and time scales in the
FA and BBL models, we note that the equilibrium spacing 
between defects sets the dynamical correlation length $\xi$
(for these one-dimensional models).  We define
the persistence time $\tau_\mathrm{p}$ by 
$\langle p_i(\tau_\mathrm{p}) \rangle=1/\mathrm{e}$, where
$p_i(t)$ is the persistence function, defined above.  As
in the pure FA model, $\tau_\mathrm{p}$ scales with the 
time taken for an excitation to propagate a distance $\xi$,
so we identify $\tau_\mathrm{p}\sim \diff^{-1} \xi^z$,
consistent with (\ref{equ:chi4_scaling}).  
To obtain the scaling of $\diff$ with the excitation 
density $\eta$ or $c$, it is useful to
rephrase the scaling argument associated
with the effective dynamics.  In the BBL model,
Eq.~(\ref{equ:rate_dist}) implies
that the fraction of sites with rate $r_i<r$ scales 
as
\begin{equation}
n(r)\sim (r/\diff_0)^{1/\mu}
\label{equ:n_r}
\end{equation}
for small $r$.  Thus, moving a distance
$\ell$ typically requires the particle to cross a barrier
whose transmission rate is $r_i\simeq \diff_0\ell^{-\mu}$. 
The time taken to cross such a barrier is typically 
$\tau_i\sim \ell/r_i\sim \diff_0^{-1} \ell^{1+\mu}$.  This is
consistent with (\ref{equ:t-ell}), and it allows us to
identify the coefficient $\diff$ in (\ref{equ:chi4_scaling})
with $\diff_0$ in~(\ref{equ:rate_dist}).  In the
FA model, crossing a barrier with
transmission rate $r_i$ typically requires a spin to flip
from state $0$ to state $1$, and this process occurs with rate
$e^{-\beta}\sim c$.  Thus, moving a distance $\ell$ typically requires
the crossing of a barrier with $r_i\simeq \ell^{-\mu}$, which takes
a time $\tau_i\sim \ell/(r_i c)\sim c^{-1} \ell^{1+\mu}$.  Thus,
we identify the coefficient $\diff$ in (\ref{equ:chi4_scaling})
with the inverse excitation density $c$.
Overall, for $\mu>1$,
we arrive at $\tau_\mathrm{p} \sim c^{-2-\mu}$ for the FA model,
and $\tau_\mathrm{p} \sim [\ln(1/\eta)]^{-\mu}
\eta^{-1-\mu}$ for the BBL model, where
we used $\diff_0\sim\gamma^{-\mu}\sim 
[\ln(1/\eta)]^{-\mu}$. 

We conclude that the effective dynamics scheme
presented here captures the propagation of
mobility excitations in the BBL and disordered FA models on
large length and time scales,
even though the BBL model has non-trivial dynamical correlations in
the densities $\rho_i$ which the coarse-grained FA model neglects.
This analysis demonstrates that the scaling properties of the 
persistence time and the
four-point susceptibility, as expressed in (\ref{equ:exponents_pure}),
(\ref{equ:exponents_dis}) and 
(\ref{equ:chi4_scaling}), can be understood in terms
of independently propagating (non-cooperative) excitations.

\subsection{Long-time limit}
\label{sec:long-time}

So far, we have considered time scales up to the persistence time $\taup$: 
excitations move distances smaller than their typical spacing, 
and can be treated independently.
We now turn to much longer time scales.  The
assumption of independently propagating defects in one
dimension leads to a persistence function consistent
with the results of~\cite{BBL05}: 
\begin{equation} \label{equ:stretch}
p(t) \equiv \langle p_i(t) \rangle = \exp\left[-\frac{(\Omega t)^{1/z}}{\xi}\right]
\end{equation}
for $t\gg 1$.

However, for times larger than $\taup$, this prediction fails.
For example, in the site-disordered FA model, the fraction
of sites with rate $r_i < r$ is $n(r) = r^{1/\mu}$.  
At infinite temperature ($\beta=0, c=1/2$), the facilitation
constraint in the FA model can be ignored (most sites are
faciliated).  In that case, the typical
time taken to flip (for the first time) a site with initial
rate $r_i$ is $\tau_i=1/r_i$.  Thus, the
persistence function decays as $\langle p_i(t) \rangle_{\beta=0}
\simeq n(t^{-1}) \simeq t^{-1/\mu}$.   
Lowering the temperature in
the FA model only slows down the dynamics, so 
$\langle p_i(t) \rangle \gtrsim t^{-1/\mu}$ for all times and
temperatures.  Thus, (\ref{equ:stretch}) must break down
at long times: we attribute this breakdown to
the fact a single site with a small rate $r_i$ can block
the motion of several excitations.

We now consider this long-time regime in more detail,
and return to the effective dynamics picture, working
with a finite density of excitations, $\eta$.
If the density of relevant barriers is larger than the
density of excitations,  each effective trap typically
contains at most one excitation, and excitations can
be treated independently.   However, when the spacing between relevant
barriers becomes larger than the distance between excitations, one enters
a different regime.  To see this, note that
the typical time scale associated with rearrangement of a
`slow' (relevant) site $i$ in the 
BBL model is generically $\tau_i = 1/[r_i(\eta_{i-1}+\eta_{i+1})]$ 
where $\eta_{i-1}$ and $\eta_{i+1}$ are the excitation densities 
in the effective 
traps to left and right of site $i$.  (Recall that $r_i$ 
is the rate with which the relevant site rearranges, 
given that there is an excitation adjacent to that
site.  Thus, the time taken to flip a relevant site depends on
the density of excitations in the adjacent traps.)

In the short-time regime where there are many
more effective traps than there are excitations, then we can
write $\eta_i\simeq 1/l_i$ if trap $i$ contains an excitation, and 
$\eta_i=0$ otherwise (as
above, $l_i$ is the width of effective trap $i$).  Considering
a site $i$ for which one of the adjacent traps contains an excitation,
we arrive at the scaling relation $\tau_i \simeq l_i/r_i \simeq \ell/r_i$,
as discussed above.
However, if there are more excitations than effective traps,
we expect the density in each trap to be close to its equilibrium
value $\eta_i\simeq \eta_{i+1}\simeq \eta$.  Thus, we expect 
$\tau_i\sim (2 \eta r_i)^{-1}$ for $\ell\eta\gg 1$.
In both cases,
for a given time $t$, we use $n(r)\simeq (r/\diff)^{1/\mu}$
to evaluate the fraction of sites with
$\tau_i<t$: the mean spacing between these ``relevant'' 
sites is $\ell(t)$.  The result is
\begin{equation} \label{equ:ell-scaling}
\ell(t) \sim \left\{ \begin{array}{ll} (\diff t)^{\frac{1}{1+\mu}}, 
   & \ell(t)\ll\xi \\
                                       (\diff t/\xi)^{1/\mu},  
   & \ell(t)\gg\xi
 \end{array} \right.
\end{equation} 
where we have used $\xi\sim\eta^{-1}$.  
For long times, the exponent
$\frac 1\mu$ sets the time dependence of $\ell(t)$: this
result applies for all $\mu>0$.  In the short time regime,
(\ref{equ:ell-scaling}) is consistent with the analysis of the
previous section, and with Eq.~(\ref{equ:chi4_scaling}),
as long as $\mu>1$.  
However, if $\mu<1$, the motion of excitations is
diffusive: thus, in the short time regime, 
there is no distinction between relevant and irrelevant
barriers. 
This means that if $\mu<1$, the spacing between relevant
barriers, $\ell(t)$, is only well-defined in the long time limit,
and the short-time scaling regime of (\ref{equ:ell-scaling}) does
not exist. 

We observe that in the long time
limit, $\ell(t)$ represents the mean spacing between
isolated sites with small rates $r_i$, and these sites
dominate the long-time limit of the persistence function.
That is, in the long time scaling regime, (\ref{equ:stretch})
is replaced by
\begin{equation} \label{equ:pers-long}
p(t)\simeq \ell(t)^{-1} \simeq (\Omega t/\xi)^{-1/\mu}.
\end{equation}  

\begin{figure} 
\epsfig{file=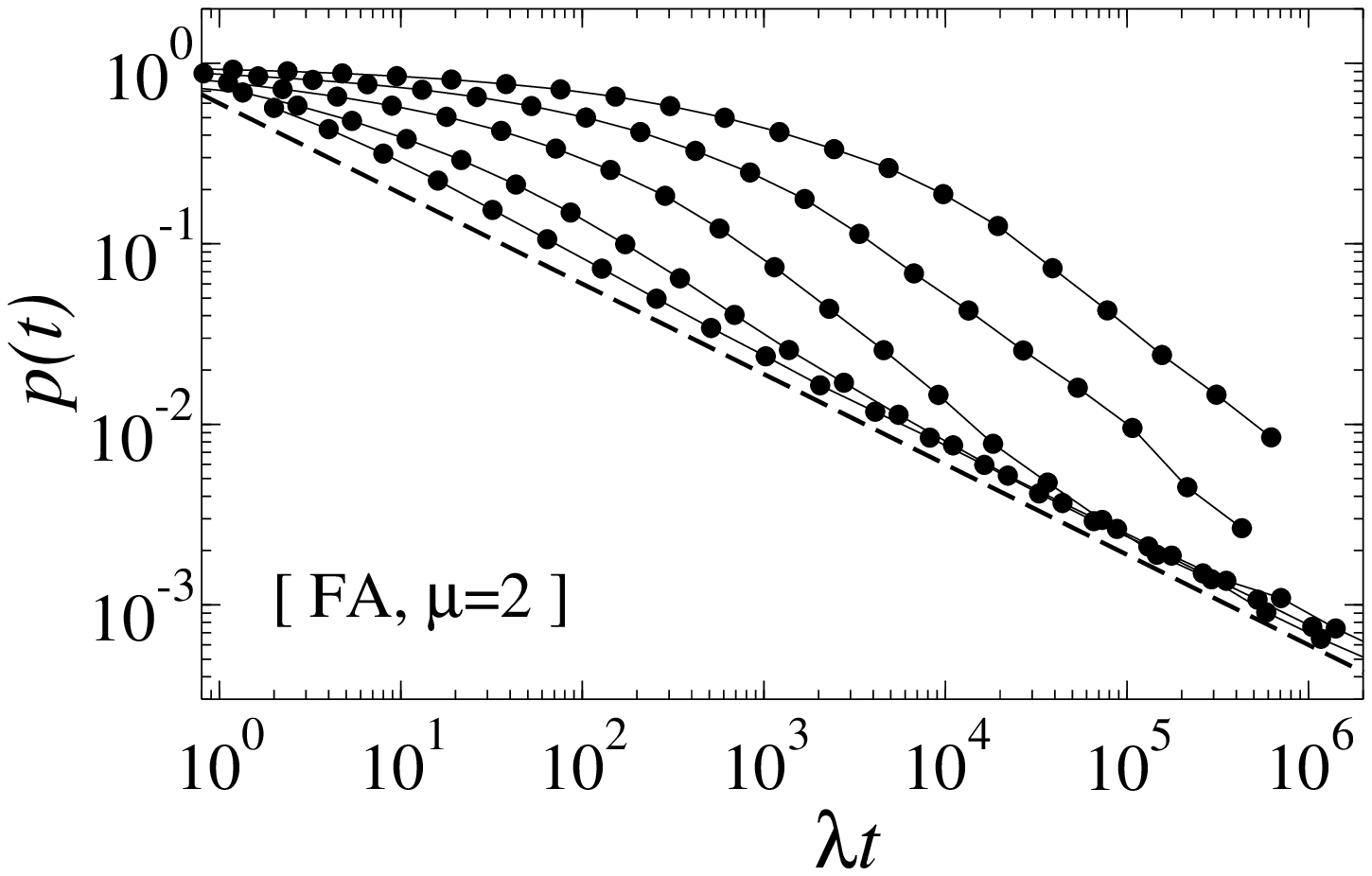,width=7.6cm}
\epsfig{file=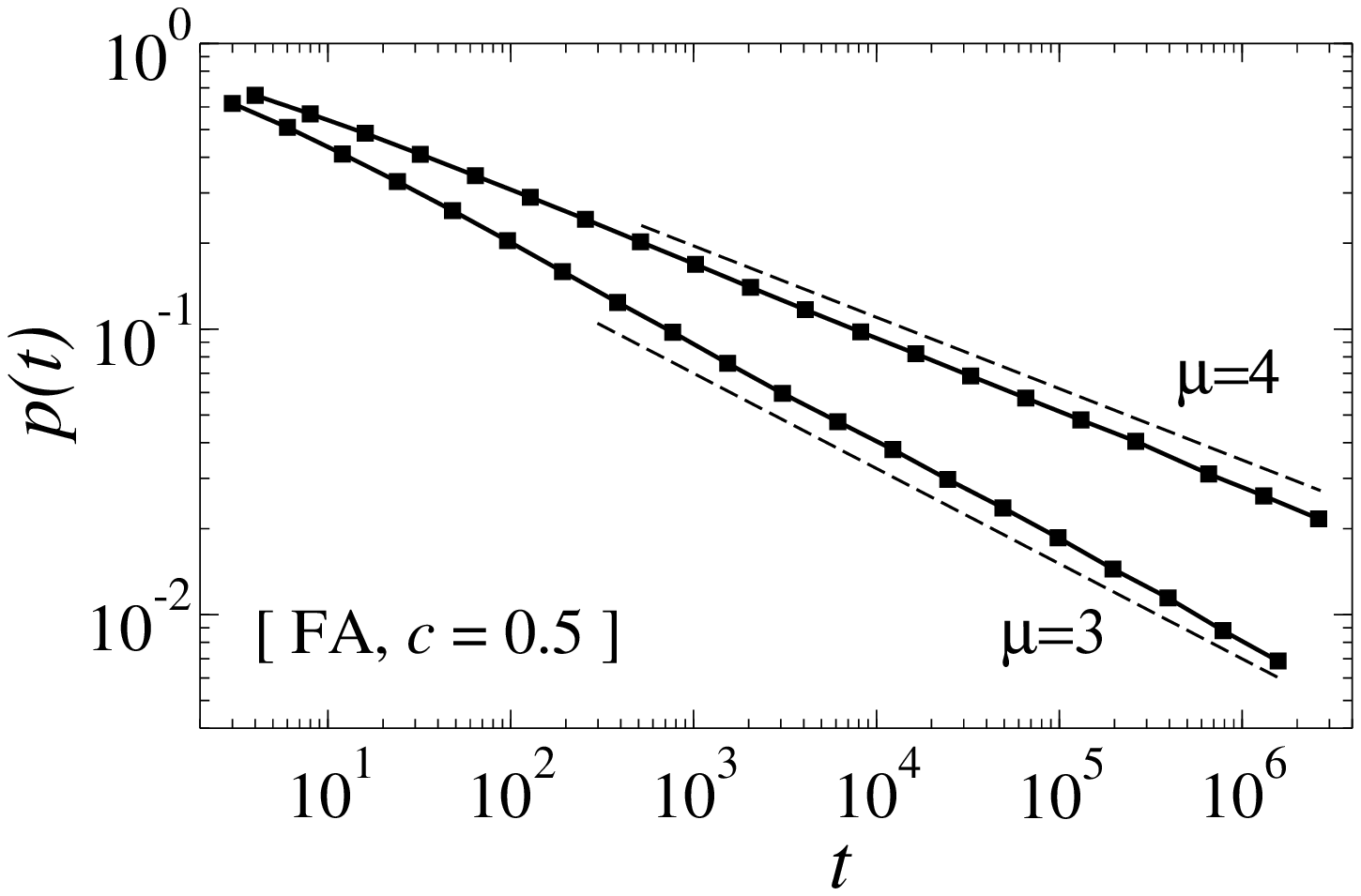,width=7.6cm}
\caption{
(Top) Persistence function, plotted against the scaling
variable
$\lambda t$ in the site-disordered FA model.
We show data for 
that $c=0.5,0.27,0.12,0.047,0.018$ (decreasing from left to right,
these are inverse temperatures $\beta=0,1,2,3,4$).
As discussed in the text, the persistence time scales as 
$\taup\sim c^{-2-\mu}\sim c^{-4}$,
but we use the variable variable $\lambda t$ to verify
the long time prediction of (\ref{equ:pers-ell}):
the dashe lines shows a power law with exponent $\frac 12$.
(Bottom) Persistence at
infinite temperature and varying $\mu$.  Again, dashed lines
show power laws with exponents predicted by (\ref{equ:pers-ell}).
}
\label{fig:pers_coll}
\end{figure}

The crossover between the two
scaling regimes occurs when
$\ell(t)\simeq \xi \simeq \eta^{-1}$ or $c^{-1}$, respectively.  
This can be observed in
the long-time behaviour of the persistence function in the 
site-disordered FA model.  
To obtain the long-time limit of this function more quantitatively,
we decompose the persistence $p(t)=cp_1(t)+(1-c)p_0(t)$ 
into contributions from sites that were initially
in states $1$ and $0$ [these two populations have
weights $c$ and $(1-c)$ respectively].  Then, in the long time regime
$\ell\gg\xi$, we have $\tau_i\simeq (2c r_i)^{-1}$ for sites
that were initially in state 1, given that each of the two
facilitating neighbour sites contains a defect with probability $c$.
For those sites initially  
in state 0, the spin flip rate is suppressed by 
$\ee^{-\beta}$, leading to $\tau_i\simeq (2c\ee^{-\beta}r_i)^{-1}$.
The persistence functions are then estimated as the density of sites
with $\tau_i>t$, giving $p_1(t)=n(1/(2ct))$ and
$p_0(t)=n(1/(2c\ee^{-\beta}t))$, respectively.
Using $n(r)=r^{1/\mu}$, we thus arrive at
\begin{equation} \label{equ:pers-ell}
 p(t) \simeq (\lambda t)^{-1/\mu} 
\qquad \ell(t) \gg \xi \sim c^{-1}
\end{equation}
with 
\begin{equation}
\lambda = 2c\ee^{-\beta}[1 + c(\ee^{-\beta/\mu}-1)]^{-\mu}
\end{equation}
In the limit of dilute excitations ($c\ll 1$) 
this reduces to $\lambda\sim c^2$.  Identifying 
$p(t)$ with $\ell(t)^{-1}$, this result is consistent
with (\ref{equ:ell-scaling}), since we argued in 
Section~\ref{sec:eff} that $\diff\sim c$ for the FA model.

Results are shown in Fig.~\ref{fig:pers_coll}: at
infinite temperature the power-law behaviour of the
persistence is clear.  
At lower temperatures, the crossover
to power-law behaviour occurs deep in the tails of
the persistence [$p(t)\lesssim c$].  
In the FA model,
this long time regime 
can be demonstrated by simulations 
at high temperature, on relatively short time scales.  
However, in the BBL model, the equivalent
of the high-temperature regime requires small $\gamma$,
increasing the prefactor $\diff_0$, and reducing the 
fraction of sites with small $r_i$.
This means that very long simulations are required to access 
the long-time limit in the BBL model,
and we do not show numerical data in this case.
However, the simulations of the 
site-disordered FA model confirm the validity of the
arguments of this section, which apply
to both FA and BBL models.  
(To observe the long-time
regime in the bond-disordered FA model, one would need to define
and measure a persistence observable on bond $i$, associated
with the rearrangement of density across that bond.)

\section{Probe particles}
\label{sec:probes}

It is a familiar feature of kinetically constrained models
that propagation of probe particles is different from that of
excitations \cite{JungGC04,BGC-Fick,Chandler06,Kelsey08}.
We now turn to probe particle motion in the BBL and disordered
FA models.

\subsection{Probes in the BBL model}

We introduce (non-interacting) 
probe particles to the BBL model as follows. A probe can
move along a bond when density rearranges across that
bond. If the bond connects sites $i$ and $i+1$, then after
the rearrangement, the probe occupies site $i$
with probability $\rho_i/(\rho_i+\rho_{i+1})$. The joint stationary
distribution for the probe position and the BBL densities is 
\begin{equation} \label{equ:probe_dist}
P_\mathrm{probe,stat}(X,\{\rho_i\}) \propto 
 \rho_{i=X} \prod_i P_\mathrm{s}(\rho_i),
\end{equation}
where $X$ is the position (site index) of the probe. Thus,
the probability of finding a probe on site $i=X$ is  
is proportional to the local BBL density on that
site, $\rho_{i=X}$. This is 
consistent with the probe representing a typical particle
in the BBL model, before the coarse-graining into the densities
$\rho_i$ is carried out.   An alternative rule, which is more 
consistent with the effective FA model described below, is 
to assign a probe with equal probability to sites $i$ and $i+1$.
As we discuss below, the excitation motion in these models
sets bounds on the motion of the probes: we are primarily concerned
with the situation in which these bounds are saturated, in
which case details of the microscopic probe motion should
be irrelevant.  When the bounds are not saturated,
the choice of dynamical rule does produce quantitative differences,
although qualitative features are preserved.

\subsection{Probes in the FA model}

We couple probe particles to the FA model using the method
of \cite{JungGC04}. Probes can hop between pairs of adjacent sites 
only when both sites have $n=1$; 
they attempt these hops with unit rate.  With these rules, the equilibrium
distribution analogous to~(\ref{equ:probe_dist}) is independent of $X$:
that is, the distribution of the probe position decouples from the
excitation variables $n_i$ and the rates $r_i$.  

From the data presented above on the excitation dynamics of the
site-disordered and bond-disordered FA models, one might expect that
the two model variants also exhibit similar probe dynamics.  However,
this is not the case because barriers to excitation diffusion act
differently on the probes. To see this, consider the site-disordered
model, and suppose that site $i$ starts with $n_i=0$ and with a small
rate $r_i$.  The probe cannot cross this site until its excitation
state changes to $n_i=1$.  The rate for this is of order $c r_i$, and
so the rate for a probe to cross this site also vanishes with $r_i$: high
barriers for excitations (small $r_i$) are also high barriers for probes in the
site-disordered FA model.

Now consider the bond-disordered FA model, 
focusing on a particular bond
$i$, with a small rate $r_i$.  The probe particle can
cross this bond if $n_i=1$ and $n_{i+1}=1$: this state can occur
on time scales much shorter than $(cr_i)^{-1}$ if an excitation
arriving from the right facilitates $n_{i+1}$ and another excitation
arriving from the left facilitates $n_i$. This process sets a rate for
crossing the slow bond that is independent of $r_i$. So the barriers
for excitation diffusion have a much smaller effect on probe
propagation in the bond-disordered model. (One way to avoid this
behaviour would be to allow the probe to move along bond $i$ only when
the rate for that bond is randomised, but we have not pursued this as
we wanted to keep the probe dynamics similar to that used
in~\cite{JungGC04}.)

It is clear that in the original BBL model, the barriers for
excitation diffusion do also act on probes. A high barrier here is a
site with density $\rho_i\approx 2$. This can take part in a
rearrangement only once a rearrangement of neighbouring sites has
produced a low-density $\rho_{i-1}<2-\rho_i$ or $\rho_{i+1}<2-\rho_i$.
The rate for these processes, and hence for probe diffusion across
site $i$, vanishes as $\rho_i\to 2$. In summary, only the
site-disordered FA model can provide an accurate representation of the
BBL probe dynamics because it retains the effect of high barriers on
the probes. We therefore do not consider the bond-disordered case in
the following.

\begin{figure}
\epsfig{file=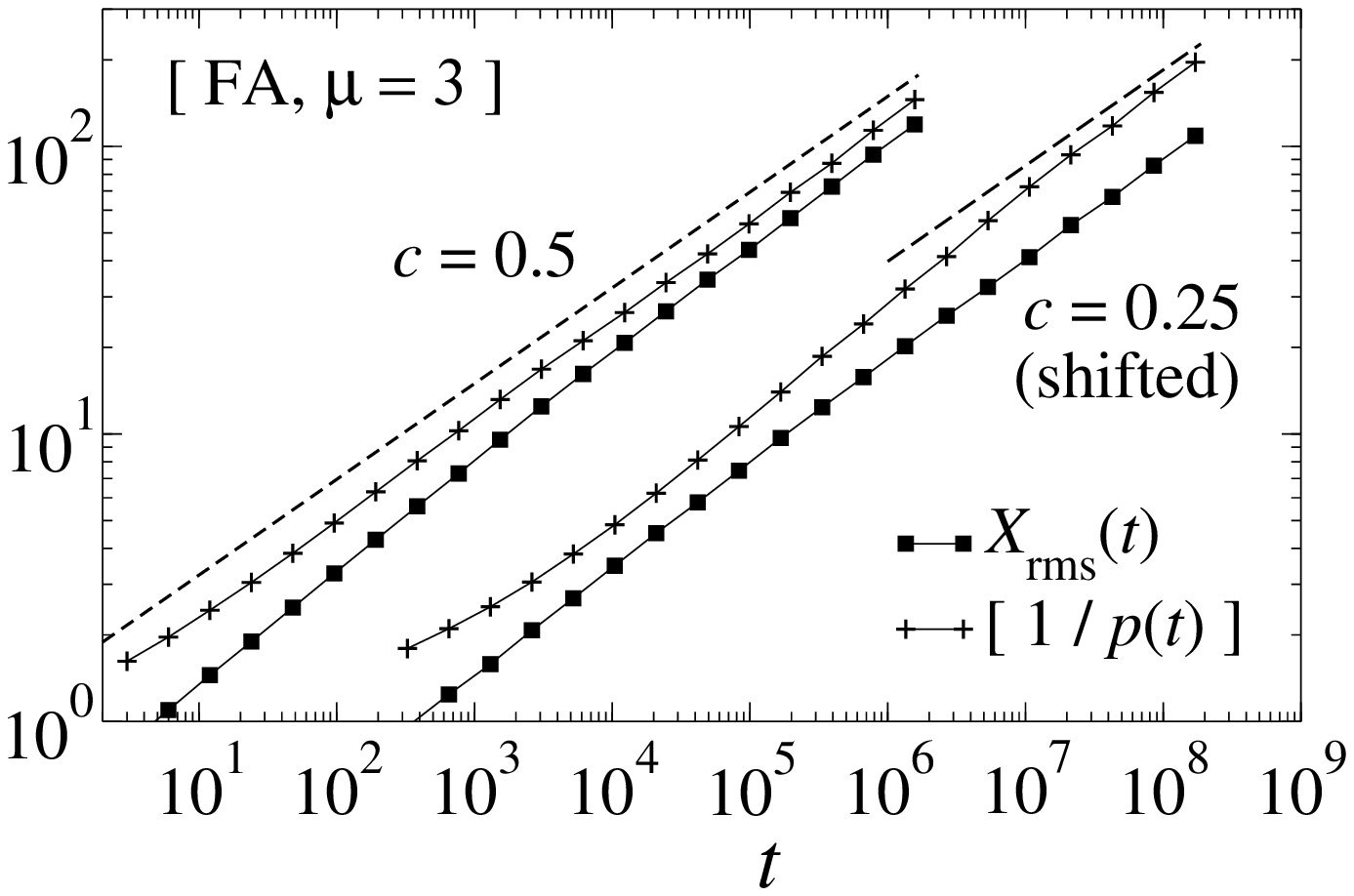,width=7.6cm}
\epsfig{file=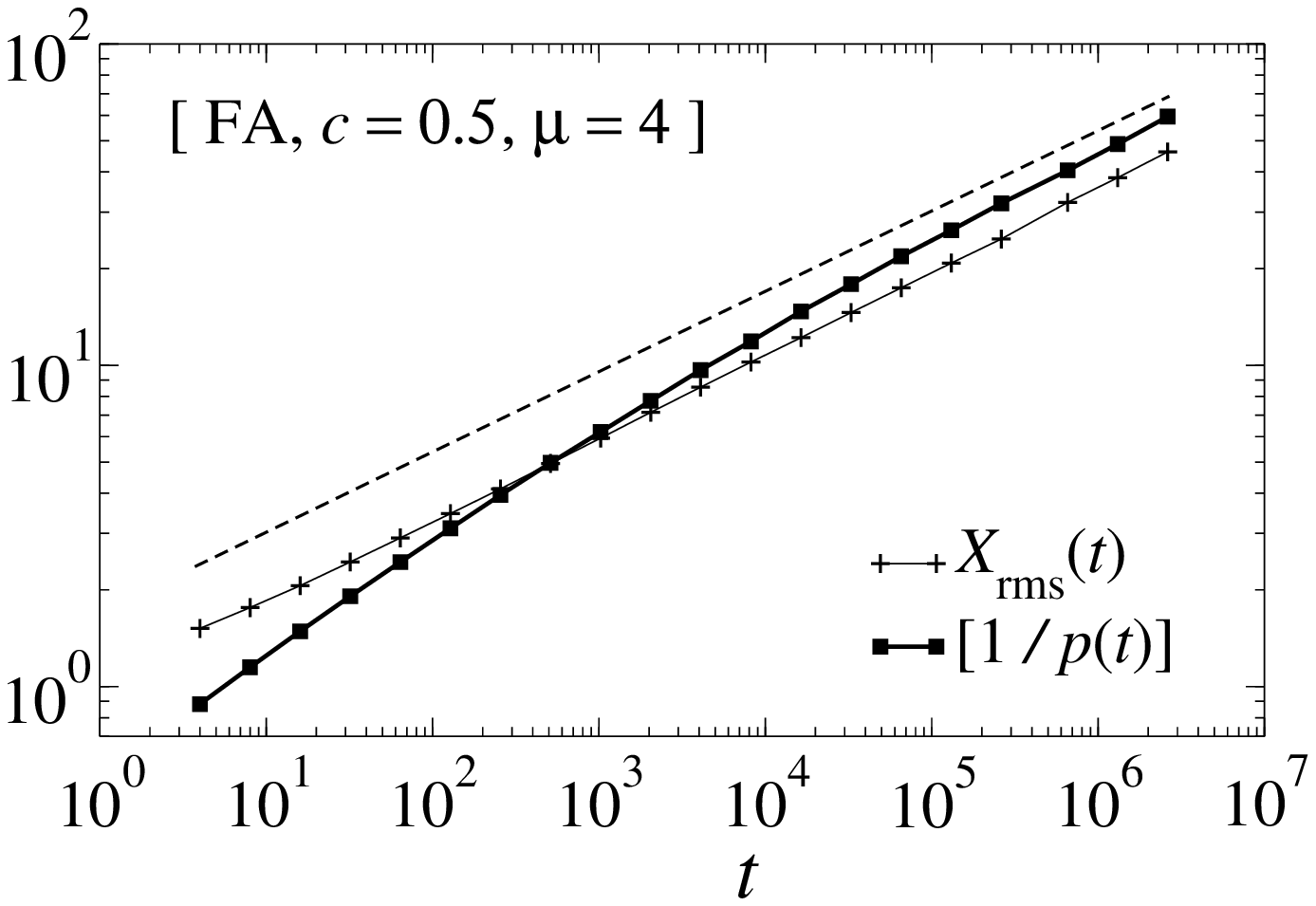,width=7.6cm}
\caption{Comparison of root mean square probe displacement 
$X_\mathrm{rms}(t)= \langle X(t)^2 \rangle^{1/2}$ with the inverse 
of the persistence function $1/p(t)$, for 
the site-disordered FA model.  We show
data at $\mu=3$ (top panel) and $\mu=4$ (bottom panel).
We have offset the data at $c=0.25$
for clarity: the ordinate is $10t$ in that case.
In the long-time limit, 
the persistence scales as $t^{-1/\mu}$ (this power law
is illustrated with dashed lines).  At long times the persistence 
sets a bound on the scaling of the mean square displacement,
which appears to saturate for long times.  
}
\label{fig:fa_probe_b0}
\end{figure}

\subsection{Results for probe motion}

In the BBL and site-disordered FA models, the preceding
discussion illustrates that sites with small rate $r_i$ are able
to block the propagation of probes.  Taking the FA model
for concreteness, the probe cannot pass any site for
which $n_i=0$ for all times between $0$ and $t$.  As discussed 
in Section~\ref{sec:long-time}, the mean spacing between
these sites scales as $t^{1/\mu}$ for large times $t$.  This
sets a limit on probe motion
\begin{equation} \label{equ:probe_scaling}
\langle X(t)^2 \rangle \lesssim \ell(t)^2 \sim t^{2/\mu}
\end{equation}
For $\mu<2$, this bound is irrelevant: the probes simply
diffuse.  However, for $\mu>2$, we expect
this bound to be saturated at large times.
At long times, we have $\ell(t)\sim p(t)^{-1}$,
[recall (\ref{equ:pers-long})].  Thus, 
Fig.~\ref{fig:fa_probe_b0} demonstrates that the bound
(\ref{equ:probe_scaling}) does saturate at long times,
although we note that the times required are quite large,
even at infinite temperature ($c=1/2$).  
Physically, the length scale $\ell(t)$ represents
the size of an effective trap: saturation of the bound
requires that the probe particle explores the whole
of the trap before the barriers delimiting the trap
become irrelevant.  
The scaling arguments presented here do not allow us to
estimate the time required to reach this regime.
However, Eq.~(\ref{equ:probe_scaling}) shows that probe
propagation must be asympotically subdiffusive for
all $\mu>2$, and the data are consistent
with saturation of this bound throughout this regime.

We emphasise that while Fig.~\ref{fig:fa_probe_b0}
demonstrates that (\ref{equ:probe_scaling}) holds on long
time scales in the FA model, the scaling arguments presented
here apply equally well to the BBL model, so asymptotic
probe motion in that model
must be subdiffusive for $\mu>2$.

\begin{figure}
\epsfig{file=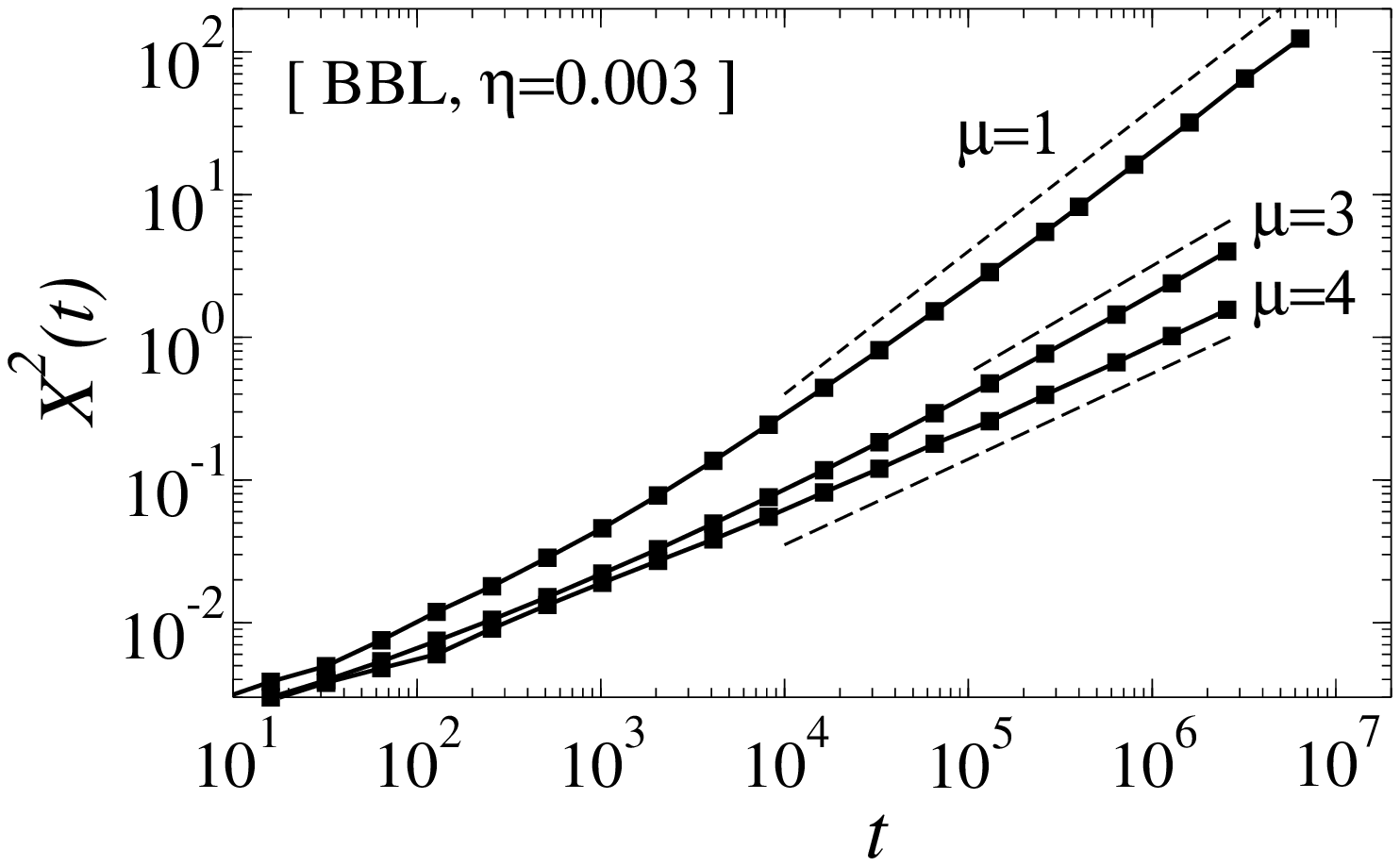,width=7.6cm}
\epsfig{file=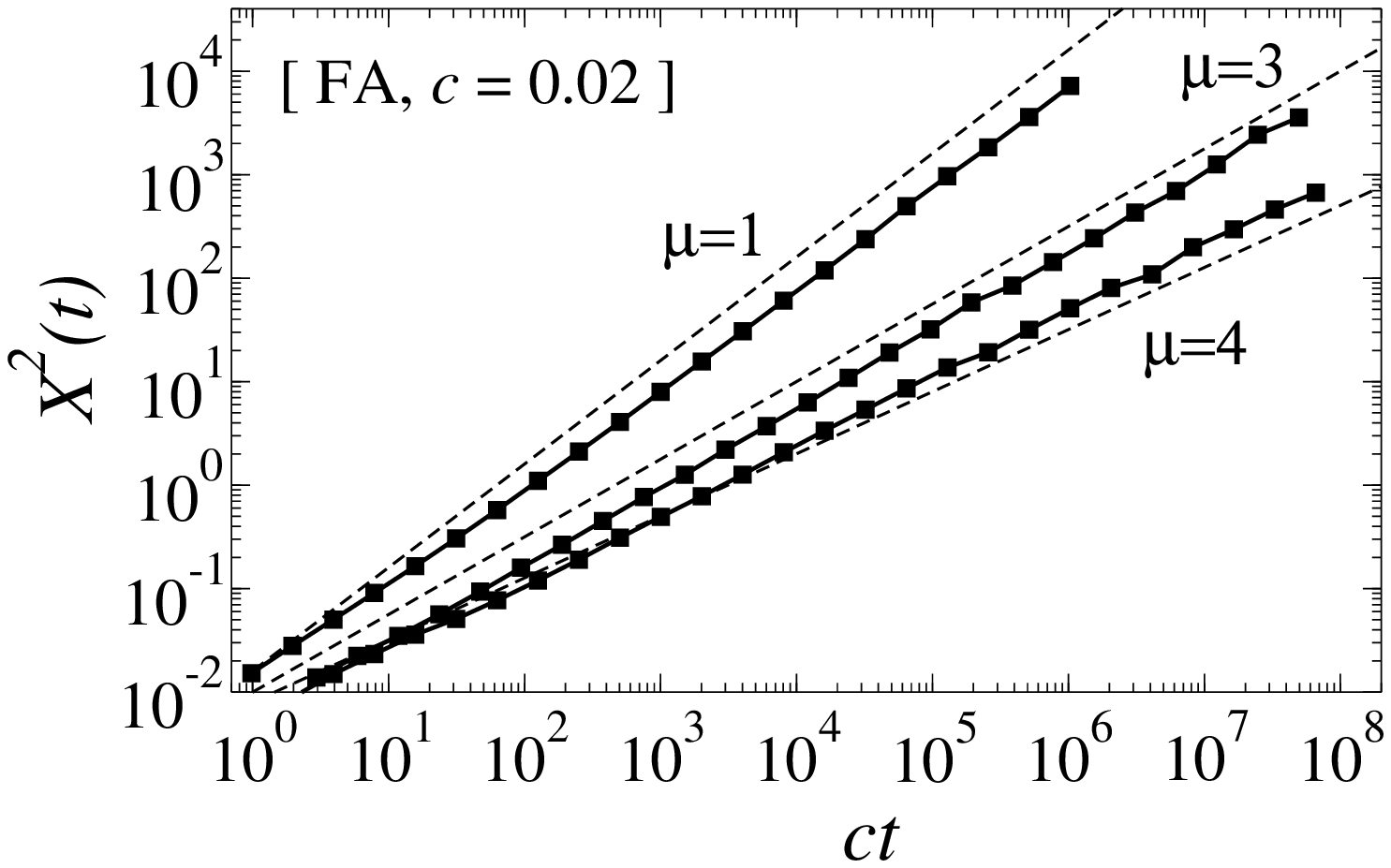,width=7.6cm}
\caption{Mean-square probe displacements in the BBL and 
FA models, for times shorter than $\taup$. 
(Top) BBL model.  For $\mu>2$, the dashed lines show the predictions
of (\ref{equ:probe_short_scaling}); for $\mu=1$, we show
a diffusive law $\langle X^2(t)\rangle \sim t$.
(Bottom) Site-disordered FA model, with the same power-law
predictions.  In both cases, we expect a crossover at
long times to the asymptotic scaling of 
(\ref{equ:probe_scaling}). 
}
\label{fig:probe}
\end{figure}

We now turn to time scales shorter than $\taup$, for
which the length scale $\ell$ again
sets a bound on probe motion.  Following~\cite{BGC-Fick}, 
we decompose the probes into two populations: those
that have moved at least once, and
those which have not moved at all.  We denote
the fraction of probes that have moved at least
once by $\pi(t)$.  Confinement
of probes by sites that are persistently unfaciliated 
again sets an upper bound on the displacement of probes:
\begin{equation} \label{equ:probe_short_bound}
\frac{\langle |X(t)|^n \rangle}{1-\pi(t)} \lesssim \ell(t)^n
\end{equation}
where the left hand side is the $n$th moment of the distance
moved by those probes that have moved at least once,
and the scaling of $\ell(t)$ was given in (\ref{equ:ell-scaling}).
Again, saturation of this bound occurs when motion of the probe
particle within the effective trap is fast enough 
that the the probe can delocalise within the trap
before the barriers delimiting the trap become irrelevant.
In the joint limit of large time and large $\mu$, 
the time scales associated with
adjacent barriers become 
well-separated~\cite{DoussalMF99,Jack-duality-08}, allowing
equilibration to take place. 
Thus, for large $\mu$, we expect the bound of~(\ref{equ:probe_short_bound})
to be saturated for times $t\gg 1$, even if $t\ll \taup$.  

Assuming saturation of the bound (\ref{equ:probe_short_bound})
and $t\ll\taup$, the probe
persistence scales as $1-\pi(t)\sim 1-\exp(-\ell(t)/\xi) \sim
\ell(t)/\xi.$
(This is the same scaling as for the excitation persistence in
(\ref{equ:stretch}).) 
 Combining this with (\ref{equ:ell-scaling}), we arrive at
\begin{equation} \label{equ:probe_short_scaling}
\langle |X(t)|^n \rangle \sim \xi^{-1} (\Omega t)^{\frac{1+n}{1+\mu}},
\label{equ:nth-moment}
\end{equation}
Our simulations are restricted to finite
time scales and values of $\mu$ that are not very large,
so we are not able to investigate this bound in detail.
However, the results shown in Fig.~\ref{fig:probe} are certainly
consistent with the prediction 
of~(\ref{equ:probe_short_scaling}).

\section{Conclusion}

To summarise our main results,
we have established that probes and mobility excitations
both propagate subdiffusively in the BBL model, and that this subdiffusive
behaviour can be reproduced in a simple effective FA model. This
observation allows us to analyse the subdiffusive motion, and to predict
the dynamical exponents for both excitations and probes in the subdiffusive
regime [Eqs.~(\ref{equ:exponents_dis}) and (\ref{equ:probe_scaling})].
A key part of the reasoning consists in showing that quenched and
annealed disorder lead to qualitatively the same behaviour.
This allowed us to deduce that
correlation length and time scales are related in these
models, by $\tau\sim\xi^{1+\mu}$.  When $\mu$ is large,
we conclude that the very broad distribution of
rates in these models leads to a
relaxation time that increases much more quickly than
the associated length scales, on approaching the glass
transition.  

We also identify two kinds of subdiffusive motion in these models.
On time scales $1\ll t\ll \taup$, mobility excitations propagate
independently and subdiffusively, according to
\begin{equation} \label{equ:conc_x}
\langle X^2 \rangle \simeq (\Omega t)^{\frac{2}{1+\mu}}
\end{equation}
One has to remember though that this $\langle X^2\rangle^{1/2}$ does
not define a lengthscale for probe motion, since it arises from an
average over a dominant population of probes that have not yet moved,
and a smaller population that has moved by $(\Omega
t)^{1/(1+\mu)}$. For the same reason the exponent for the scaling of
$\langle |X(t)|^n\rangle$ in Eq.~(\ref{equ:nth-moment}) is not simply
proportional to $n$.
On time scales $t\gg\taup$, excitations coagulate and branch,
and it is not consistent to discuss motion of a single excitation.
However, in this long-time regime, probe particles propagate
subdiffusively, according to
\begin{equation} \label{equ:conc_X}
\langle X^2 \rangle \simeq (\Omega t/\xi)^{2/\mu}
\end{equation}
The presence of different dynamical exponents for probes
and excitations may seem surprising, but we
emphasise that (\ref{equ:conc_x}) and (\ref{equ:conc_X})
apply in separate scaling regimes. (When the concentration of
excitations is small, 
the persistence time $\taup\gg 1$ separates two well-defined
scaling regimes; of course $t$ is always taken to be large compared
to unity.)

Conceptually, it is interesting to note that in the pure FA model at
low temperature, relaxation is controlled by rare active sites
(defects). In the disordered model, on the other hand, rare inactive
regions (sites with small $r_i$) play at least as important a role.

Finally, our results for probe particles imply that 
the Stokes-Einstein relation~\cite{DHReviews} 
between relaxation time and probe diffusion constant, $D\tau\sim1$, 
has broken down completely in these systems. In the pure FA model,
$D\tau$ diverges at low temperatures~\cite{JungGC04}.  On the
other hand, in the disordered model, the presence of sites
(or barriers) with arbitrarily small rate $r_i$ means that
the persistence decays as a power law for large times,
while the motion of the probes is subdiffusive even
in the long time limit.  However, we can define
an analogue of the Fickian length 
$\ell_\mathrm{F} = \langle X^2(\taup) \rangle^{1/2}$ which
represents the distance travelled by a probe, through
repeated encounters with a single excitation~\cite{BGC-Fick}.  
If the bound of
(\ref{equ:probe_short_bound}) is saturated we arrive at
$\ell_\mathrm{F}\simeq \ell(\taup) \simeq \xi$.
For the site-disordered FA model, this leads to
$\ell_\mathrm{F}\sim c^{-1}$, at least for large $\mu$;
on the other hand,
in the pure FA model, $\ell_\mathrm{F}\sim c^{-1/2}$.
Physically, confinement of the
excitation in an effective trap means that it 
facilitates any probes in that trap very many times, 
allowing the probe to delocalise thoughout the trap.
In this way, the presence of large barriers to excitation diffusion
in the BBL and disordered FA
models strengthens the effects discussed in~\cite{JungGC04,BGC-Fick},
in which the square of the Fickian length represents the number
of hops that a probe makes through multiple
encounters with a single excitation.  

\begin{acknowledgments}
We thank L.\ Berthier, J.-P.\ Bouchaud, D.\ Chandler and
J.\ P.\ Garrahan for discussions.
While at Berkeley, RLJ was funded initially by NSF grant CHE-0543158
and later by the Office of Naval Research 
Grant No.~N00014-07-1-0689.
\end{acknowledgments}



\begin{thebibliography}{99}

\expandafter\ifx\csname natexlab\endcsname\relax\def\natexlab#1{#1}\fi
\expandafter\ifx\csname bibnamefont\endcsname\relax
  \def\bibnamefont#1{#1}\fi
\expandafter\ifx\csname bibfnamefont\endcsname\relax
  \def\bibfnamefont#1{#1}\fi
\expandafter\ifx\csname citenamefont\endcsname\relax
  \def\citenamefont#1{#1}\fi
\expandafter\ifx\csname url\endcsname\relax
  \def\url#1{\texttt{#1}}\fi
\expandafter\ifx\csname urlprefix\endcsname\relax\def\urlprefix{URL }\fi
\providecommand{\bibinfo}[2]{#2}
\providecommand{\eprint}[2][]{\url{#2}}

\bibitem{RitortS03}
For a review, see
\bibinfo{author}{\bibfnamefont{F.}~\bibnamefont{Ritort}} \bibnamefont{and}
  \bibinfo{author}{\bibfnamefont{P.}~\bibnamefont{Sollich}},
  \bibinfo{journal}{Adv. Phys.} \textbf{\bibinfo{volume}{52}},
  \bibinfo{pages}{219} (\bibinfo{year}{2003}).

\bibitem{GarrahanC02}
\bibinfo{author}{\bibfnamefont{J.~P.} \bibnamefont{Garrahan}} \bibnamefont{and}
  \bibinfo{author}{\bibfnamefont{D.}~\bibnamefont{Chandler}},
  \bibinfo{journal}{Phys. Rev. Lett.} \textbf{\bibinfo{volume}{89}},
  \bibinfo{pages}{035704} (\bibinfo{year}{2002}).

\bibitem{JungGC04}
Y.-J. Jung, J. P. Garrahan and D. Chandler, Phys. Rev. E {\bf 69}, 
061205 (2004).

\bibitem{BGC-Fick}
L. Berthier, D. Chandler and J.\ P. Garrahan, Europhys. Lett. {\bf 69}, 
320 (2005).

\bibitem{BBL05}
E.~Bertin, J.-P.~Bouchaud and F.~Lequeux, Phys. Rev. Lett. {\bf 95}, 
015702 (2005).

\bibitem{Chandler06}
D. Chandler, J. P. Garrahan, R. L. Jack, L. Maibaum and A. C. Pan,
Phys. Rev. E {\bf 74}, 051501 (2006).

\bibitem{knights}
C.~Toninelli, G.~Biroli and D.~S.~Fisher, Phys. Rev. Lett. {\bf96}, 
035702 (2006)

\bibitem{Berthier-chi4}
L.~Berthier \emph{et al.}, J. Chem. Phys {\bf126}, 184504 (2007).

\bibitem{DHReviews}
For reviews of the effects of
dynamical heterogeneity and breakdown of the Stokes-Einstein relation, 
see: H. Sillescu, J. Non-Cryst. Solids {\bf 243}, 81 (1999);
M.D. Ediger, Annu. Rev. Phys. Chem. {\bf 51}, 99 (2000). 
S.C. Glotzer, J. Non-Cryst. Solids, {\bf 274}, 342 (2000); R. Richert, J.
Phys. Condens. Matter {\bf 14}, R703 (2002); H. C. Andersen,
Proc. Natl.  Acad. Sci. U. S. A. {\bf 102}, 6686 (2005).

\bibitem{KA} 
W.~Kob and H.~C.~Andersen, Phys. Rev. E {\bf  48}, 4364 (1993);
C. Toninelli, G. Biroli and D. S. Fisher, J. Stat. Phys. {\bf 120},
  167 (2005);
C. Toninelli, G. Biroli and D.~S.~Fisher, Phys. Rev. Lett. {\bf 96}, 035702
 (2006). 

\bibitem{TLG}
J.~J\"{a}ckle and A.~Kr\"{o}nig,
  J. Phys.: Condens. Matter \textbf{6}, 7633 (1994);
A.~C. Pan, J.~P. Garrahan and D. Chandler, Phys. Rev. E {\bf 72},
  041106 (2005).

\bibitem{FredricksonA84}
\bibinfo{author}{\bibfnamefont{G.~H.} \bibnamefont{Fredrickson}}
  \bibnamefont{and} \bibinfo{author}{\bibfnamefont{H.~C.}
  \bibnamefont{Andersen}}, \bibinfo{journal}{Phys. Rev. Lett.}
  \textbf{\bibinfo{volume}{53}}, \bibinfo{pages}{1244} (\bibinfo{year}{1984}).

\bibitem{EisingerJ93}
\bibinfo{author}{\bibfnamefont{S.}~\bibnamefont{Eisinger}} \bibnamefont{and}
  \bibinfo{author}{\bibfnamefont{J.}~\bibnamefont{J\"{a}ckle}},
  \bibinfo{journal}{J. Stat. Phys.} \textbf{\bibinfo{volume}{73}},
  \bibinfo{pages}{643} (\bibinfo{year}{1993}).

\bibitem{trap}
T. Odagaki and Y. Hiwatari, Phys. Rev. A {\bf 41}, 929 (1990); 
C.~Monthus and J.-P.~Bouchaud, J. Phys. A {\bf29}, 3873 (1996).

\bibitem{Qiu03}
X.~H.~Qiu and M.~D.~Ediger, J. Phys. Chem. B {\bf 107}, 469 (2003).

\bibitem{Berthier-expt-long}
C.~Dalle-Ferrier \emph{et al.}, Phys. Rev. E {\bf 76}, 041510 (2007). 

\bibitem{Whitelam04}
S.~Whitelam, L.~Berthier and J.~P.~Garrahan, Phys. Rev. Lett. {\bf92}, 
185705 (2004);

\bibitem{FranzDPG99}
\bibinfo{author}{\bibfnamefont{S.}~\bibnamefont{Franz}},
  \bibinfo{author}{\bibfnamefont{C.}~\bibnamefont{Donati}},
  \bibinfo{author}{\bibfnamefont{G.}~\bibnamefont{Parisi}}, \bibnamefont{and}
  \bibinfo{author}{\bibfnamefont{S.~C.} \bibnamefont{Glotzer}},
  \bibinfo{journal}{Philos. Mag. B} \textbf{\bibinfo{volume}{79}},
  \bibinfo{pages}{1827} (\bibinfo{year}{1999}).

\bibitem{Berthier04}
\bibinfo{author}{\bibfnamefont{L.}~\bibnamefont{Berthier}},
  \bibinfo{journal}{Phys. Rev. E} \textbf{\bibinfo{volume}{69}},
  \bibinfo{pages}{020201} (\bibinfo{year}{2004}).


\bibitem{ToninelliWBBB05}
\bibinfo{author}{\bibfnamefont{C.}~\bibnamefont{Toninelli}},
  \bibinfo{author}{\bibfnamefont{M.}~\bibnamefont{Wyart}},
  \bibinfo{author}{\bibfnamefont{L.}~\bibnamefont{Berthier}},
  \bibinfo{author}{\bibfnamefont{G.}~\bibnamefont{Biroli}}, \bibnamefont{and}
  \bibinfo{author}{\bibfnamefont{J.-P.} \bibnamefont{Bouchaud}},
  \bibinfo{journal}{Phys. Rev. E} \textbf{\bibinfo{volume}{71}},
  \bibinfo{pages}{041505} (\bibinfo{year}{2005}).

\bibitem{JMS06}
\bibinfo{author}{\bibfnamefont{R.~L.} \bibnamefont{Jack}},
  \bibinfo{author}{\bibfnamefont{P.}~\bibnamefont{Mayer}}, \bibnamefont{and}
  \bibinfo{author}{\bibfnamefont{P.}~\bibnamefont{Sollich}},
  \bibinfo{journal}{J. Stat. Mech.} (\bibinfo{year}{2006)},
  \bibinfo{pages}{P03006}.

\bibitem{HarrowellPropensity} See, for example,
A. Widmer-Cooper and P. Harrowell, Phys. Rev. Lett. {\bf 93}, 135701 (2004);
A.~Widmer-Cooper and P.~Harrowell, J. Chem. Phys, {\bf126} 154503 (2007).

\bibitem{BouchaudG90}
J.-P. Bouchaud and A. Georges, Phys. Rep. {\bf 195}, 127 (1990)

\bibitem{JackBG05}
R. L. Jack, L. Berthier and J. P. Garrahan,  Phys. Rev. E {\bf 72}, 
016103 (2005).

\bibitem{DoussalMF99}
P. le Doussal, C. Monthus and D. S. Fisher, Phys. Rev. E {\bf 59}, 4795 (1999).

\bibitem{Jack-duality-08}
R.~L.~Jack and P.~Sollich, J. Phys. A {\bf 41}, 324001 (2008).

\bibitem{Kelsey08}
R.~L.~Jack, D.~Kelsey, J.~P.~Garrahan and D.~Chandler,
Phys. Rev. E {\bf 78}, 011506 (2008).

\bibitem{typical-foot}
As discussed in~\cite{Jack-duality-08},
these arguments based on a typical length scales $\ell(t)$ are valid
since the distribution of trap widths $l$ is much narrower than the 
distribution of time scales $\tau$, so fluctuations in $l$ can
be neglected.

\end{thebibliography}
\end{document}